\documentclass[a4paper, amsfonts, amssymb, amsmath, reprint, aps, showkeys, superscriptaddress, twoside, twocolumn, nofootinbib]{revtex4-2}
\usepackage[globalcitecopy]{bibunits}
\usepackage[switch]{lineno}
\usepackage{amsthm}
\usepackage{mathtools}
\usepackage{physics}
\usepackage{xcolor}
\usepackage{graphicx}
\usepackage[left=23mm,right=13mm,top=35mm,columnsep=15pt]{geometry}
\usepackage{adjustbox}
\usepackage{placeins}
\usepackage[T1]{fontenc}
\usepackage{lipsum}
\usepackage[english]{babel}
\usepackage[utf8]{inputenc}
\usepackage{csquotes}
\usepackage[mathscr]{eucal} 
\usepackage{bibunits} 
\usepackage[pdftex, pdftitle={Article}, pdfauthor={Author}]{hyperref} 
\usepackage[capitalize]{cleveref}
\usepackage{multirow}
\usepackage{subfig} 

\newcommand{\QA}{\mathrm{Q}_{\mathrm{A}}}
\newcommand{\QO}{\mathrm{Q}_{\mathrm{O}}}
\newcommand{\QI}{\mathrm{Q}_{\mathrm{I}}}
\newcommand{\QIone}{\mathrm{Q}_{\mathrm{I1}}}
\newcommand{\QItwo}{\mathrm{Q}_{\mathrm{I2}}}

\newcommand{\mZ}{\langle Z \rangle}
\newcommand{\mC}{\langle C \rangle}
\newcommand{\mN}{\langle N_{\mathrm{RTS}} \rangle}
\newcommand{\NRTS}{N_{\mathrm{RTS}}}
\newcommand{\psuccess}{p_{\mathrm{S}}}
\newcommand{\kzA}{\ket{0_{\mathrm{A}}}}
\newcommand{\koA}{\ket{1_{\mathrm{A}}}}
\newcommand{\kkI}{\ket{k_{\mathrm{I}}}}
\newcommand{\kzI}{\ket{0_{\mathrm{I}}}}
\newcommand{\koI}{\ket{1_{\mathrm{I}}}}
\newcommand{\kpI}{\ket{\psi_{\mathrm{I}}}}
\newcommand{\kzO}{\ket{0_{\mathrm{O}}}}
\newcommand{\kpO}{\ket{\psi_{\mathrm{O}}}}
\newcommand{\thetak}{\theta_{k}}
\newcommand{\thetakl}{\theta_{kl}}
\newcommand{\kkIo}{\ket{k_{\mathrm{I1}}}}
\newcommand{\klIt}{\ket{l_{\mathrm{I2}}}}
\newcommand{\mA}{m_{\mathrm{A}}}
\newcommand{\ZA}{Z_{\mathrm{A}}}
\newcommand{\ZIo}{Z_{\mathrm{I1}}}
\newcommand{\ZIt}{Z_{\mathrm{I2}}}
\newcommand{\XO}{Z_{\mathrm{O}}}
\newcommand{\ZO}{Z_{\mathrm{O}}}

\newcommand{\degrees}{^{\circ}}

\newcommand{\us}{\mathrm{\mu s}}
\newcommand{\MHz}{\mathrm{MHz}}
\newcommand{\GHz}{\mathrm{GHz}}
\newcommand{\Tone}{T_\mathrm{1}}
\newcommand{\Ttwoecho}{T_\mathrm{2}}
\newcommand{\Ttwostar}{T^*_\mathrm{2}}
\newcommand{\leakrate}{L_{1}}

\begin{document}

\title{Realization of a quantum neural network using repeat-until-success circuits in a superconducting quantum processor}


\author{M.~S.~Moreira}
\affiliation{QuTech, Delft University of Technology, P.O. Box 5046, 2600 GA Delft, The Netherlands}
\affiliation{Kavli Institute of Nanoscience, Delft University of Technology, P.O. Box 5046, 2600 GA Delft, The Netherlands}

\author{G.~G.~Guerreschi}
\affiliation{Intel Labs, Intel Corporation, Hillsboro, Oregon 97124, USA}

\author{W.~Vlothuizen}
\altaffiliation[Present address: ]{Qblox, Elektronicaweg 10, 2628 XG Delft, The Netherlands}
\affiliation{QuTech, Delft University of Technology, P.O. Box 5046, 2600 GA Delft, The Netherlands}
\affiliation{Netherlands Organisation for Applied Scientific Research (TNO), P.O. Box 96864, 2509 JG The Hague, The Netherlands}

\author{J.~F.~Marques}
\affiliation{QuTech, Delft University of Technology, P.O. Box 5046, 2600 GA Delft, The Netherlands}
\affiliation{Kavli Institute of Nanoscience, Delft University of Technology, P.O. Box 5046, 2600 GA Delft, The Netherlands}

\author{J.~van~Straten}
\altaffiliation[Present address: ]{Qblox, Elektronicaweg 10, 2628 XG Delft, The Netherlands}
\affiliation{QuTech, Delft University of Technology, P.O. Box 5046, 2600 GA Delft, The Netherlands}
\affiliation{Computer Engineering Lab, Delft University of Technology, Mekelweg 4, 2628 CD Delft, The Netherlands}

\author{S.~P.~Premaratne}
\affiliation{Intel Labs, Intel Corporation, Hillsboro, Oregon 97124, USA}

\author{X.~Zou}
\affiliation{Intel Labs, Intel Corporation, Hillsboro, Oregon 97124, USA}

\author{H.~Ali}
\affiliation{QuTech, Delft University of Technology, P.O. Box 5046, 2600 GA Delft, The Netherlands}
\affiliation{Kavli Institute of Nanoscience, Delft University of Technology, P.O. Box 5046, 2600 GA Delft, The Netherlands}

\author{N.~Muthusubramanian}
\affiliation{QuTech, Delft University of Technology, P.O. Box 5046, 2600 GA Delft, The Netherlands}
\affiliation{Kavli Institute of Nanoscience, Delft University of Technology, P.O. Box 5046, 2600 GA Delft, The Netherlands}

\author{C.~Zachariadis}
\altaffiliation[Present address: ]{Quantware B.V., Elektronicaweg 10, 2628 XG Delft, The Netherlands}
\affiliation{QuTech, Delft University of Technology, P.O. Box 5046, 2600 GA Delft, The Netherlands}
\affiliation{Kavli Institute of Nanoscience, Delft University of Technology, P.O. Box 5046, 2600 GA Delft, The Netherlands}

\author{J.~van~Someren}
\affiliation{QuTech, Delft University of Technology, P.O. Box 5046, 2600 GA Delft, The Netherlands}
\affiliation{Computer Engineering Lab, Delft University of Technology, Mekelweg 4, 2628 CD Delft, The Netherlands}

\author{M.~Beekman}
\affiliation{QuTech, Delft University of Technology, P.O. Box 5046, 2600 GA Delft, The Netherlands}
\affiliation{Netherlands Organisation for Applied Scientific Research (TNO), P.O. Box 96864, 2509 JG The Hague, The Netherlands}

\author{N.~Haider}
\affiliation{QuTech, Delft University of Technology, P.O. Box 5046, 2600 GA Delft, The Netherlands}
\affiliation{Netherlands Organisation for Applied Scientific Research (TNO), P.O. Box 96864, 2509 JG The Hague, The Netherlands}

\author{A.~Bruno}
\altaffiliation[Present address: ]{Quantware B.V., Elektronicaweg 10, 2628 XG Delft, The Netherlands}
\affiliation{QuTech, Delft University of Technology, P.O. Box 5046, 2600 GA Delft, The Netherlands}
\affiliation{Kavli Institute of Nanoscience, Delft University of Technology, P.O. Box 5046, 2600 GA Delft, The Netherlands}

\author{C.~G.~Almudever}
\altaffiliation[Present address: ]{Technical University of Valencia, Camino de Vera, 46022 València, Spain}
\affiliation{QuTech, Delft University of Technology, P.O. Box 5046, 2600 GA Delft, The Netherlands}
\affiliation{Computer Engineering Lab, Delft University of Technology, Mekelweg 4, 2628 CD Delft, The Netherlands}

\author{A.~Y.~Matsuura}
\affiliation{Intel Labs, Intel Corporation, Hillsboro, Oregon 97124, USA}

\author{L.~DiCarlo}
\affiliation{QuTech, Delft University of Technology, P.O. Box 5046, 2600 GA Delft, The Netherlands}
\affiliation{Kavli Institute of Nanoscience, Delft University of Technology, P.O. Box 5046, 2600 GA Delft, The Netherlands}

\date{\today}

\begin{abstract}
Artificial neural networks are becoming an integral part of digital solutions to complex problems. However, employing neural networks on quantum processors faces challenges related to the implementation of non-linear functions using quantum circuits. In this paper, we use repeat-until-success circuits enabled by real-time control-flow feedback to realize quantum neurons with non-linear activation functions. These neurons constitute elementary building blocks that can be arranged in a variety of layouts to carry out deep learning tasks quantum coherently. As an example, we construct a minimal feedforward quantum neural network capable of learning all 2-to-1-bit Boolean functions by optimization of network activation parameters within the supervised-learning paradigm. This model is shown to perform non-linear classification and effectively learns from multiple copies of a single training state consisting of the maximal superposition of all inputs.
\end{abstract}
\maketitle

\begin{bibunit}[naturemag]

\section{Introduction} \label{sec:introduction}

Deep learning is an established field with pervasive applications ranging from image classification to speech recognition~\cite{Goodfellow16}.
Among the most intriguing recent developments is the extension to the quantum regime and the search for advantage based on quantum mechanical effects~\cite{Biamonte17}.
This effort is pursued in a variety of ways, often inspired by the diversity of classical models and based on the concept of artificial neural networks.
Prior works have proposed quantum versions of perceptrons ~\cite{Schuld15, Tacchino19}, support vector machines~\cite{Rebentrost14,Havlicek19}, Boltzmann machines~\cite{Benedetti16}, autoencoders~\cite{Romero17}, and convolutional neural networks~\cite{Cong19,Herrmann21,Gong22}.
The advantage ranges from reducing the model size by exploiting the exponentially large number of amplitudes defining multi-qubit states, to speeding-up either training or inference by applying efficient quantum algorithms such as HHL~\cite{Harrow09} to solve systems of linear equations, or reducing the number of samples needed for accurate learning.

A promising implementation is based on variational quantum algorithms in which parametrized quantum circuits are used to prepare approximate solutions to the problem at hand. These solutions are then refined by classically optimizing circuit parameters~\cite{Cerezo21}.
However, fundamental questions must be answered on the parameter landscape~\cite{McClean18}, on the cost of the classical optimization loop, and on the expressive power of circuit ansatze~\cite{Abbas21}.
Encouraging results suggest that trainability is possible for quantum convolutional neural networks~\cite{Pesah21, Killoran19}. Still, it is recognized that loading training set data into a quantum machine accurately and efficiently is an unsolved problem~\cite{Aaronson15} and, although promising results~\cite{Harrow20,Zoufal19,Holmes20}, current solutions work only under specific assumptions. Nevertheless, the exponential complexity of states generated by ever larger quantum computers~\cite{Arute19} suggests that machine learning techniques will become increasingly important at directly processing large-scale quantum states~\cite{Huang22}.

It was noted in traditional machine-learning literature that non-linear activation functions for neurons are superior~\cite{Hornik89}.
To translate this observation to the design of quantum neural networks (QNNs), several methods have been proposed to break the intrinsic linearity of quantum mechanics.
These solutions range from the use of quantum measurements and dissipative quantum gates~\cite{Kak95}, to the quadratic form of the kinetic term~\cite{Behrman00}, reversible circuits~\cite{Wan17}, recurrent neural networks~\cite{Rebentrost18} and the SWAP test~\cite{Zhao19} with  phase estimation~\cite{Li20}.

\begin{figure*}
\centering
\includegraphics[width=\textwidth]{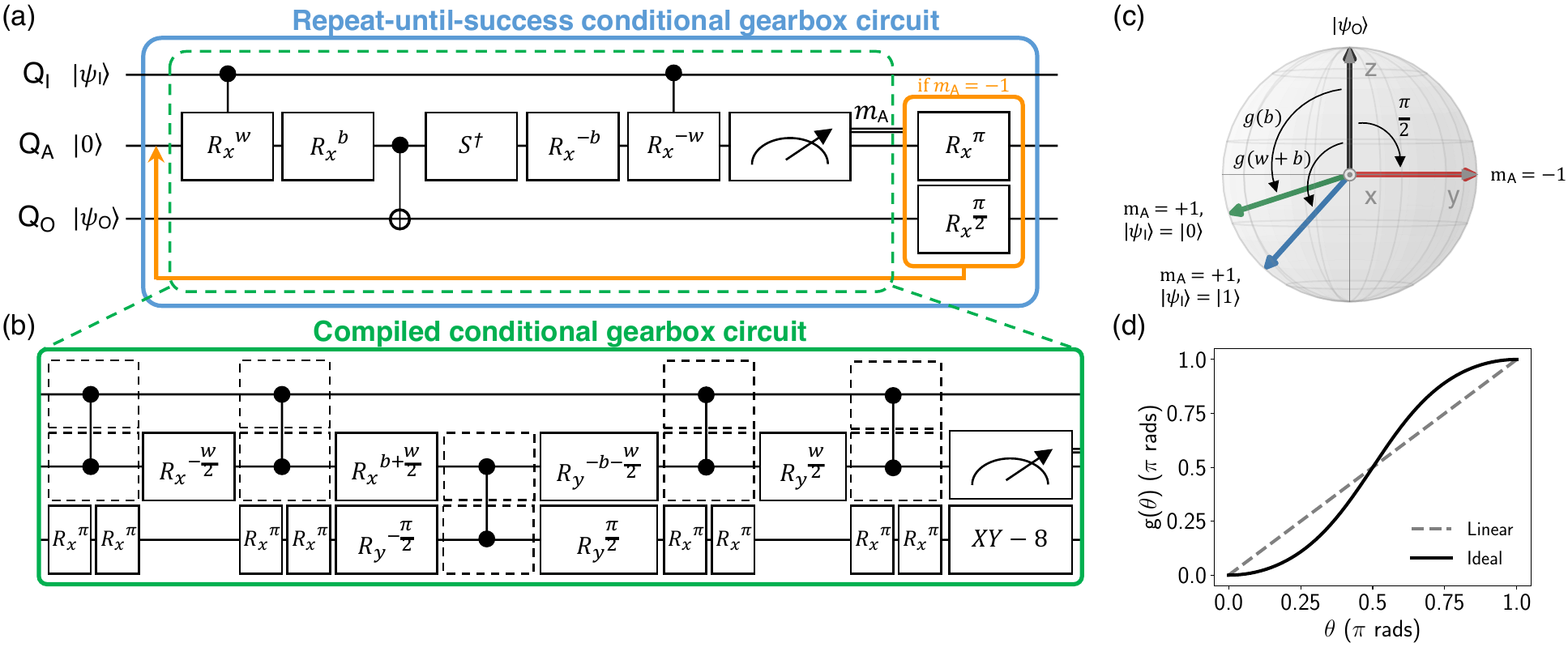}
\caption{
\textbf{Conditional gearbox circuit using repeat until success.}
(a) Three-qubit circuit with input parameters $(w,b)$ ideally implementing $R_{x}^{g(w+b)}$ on $\QO$ for $\QI=\ket{1}$ and $R_{x}^{g(b)}$ for $\QI=\ket{0}$, heralded by $\mA=+1$ (success). For $\mA=-1$ (failure), the circuit ideally implements $R_{x}^{-\frac{\pi}{2}}$ on $\QO$.
The probabilistic nature of the circuit is rectified using RUS: in case of failure, $\QA$ and $\QO$ are first reset ($R_{x}^{\pi}$ and $R_{x}^{\frac{\pi}{2}}$, respectively) and the circuit re-run. (b) Compilation into the native gate set after circuit optimization and added error mitigation (two refocusing pulses on $\QO$ during $\QI$-$\QA$ CZ gates). (c) Illustration of the ideal action of the conditional gearbox circuit on $\QO$ when starting in $\kzO$. (d) Comparison of the ideal $g(\theta)$ to a Rabi oscillation of $\QO$, showing the non-linearity of $g$.
}
\label{fig:conditional_gearbox}
\end{figure*}

Previous work in this context~\cite{Herrmann21} has shown the implementation of neural networks applied in post-processing to the classical results of measurements.
Here, we experimentally demonstrate a quantum neural network architecture based on variational repeat-until-success (RUS) circuits~\cite{Paetznick14,Bocharov15}, that is implemented in a fully coherent way, handling quantum data directly, and in which real-time feedback is used to perform the internal update of neurons. In this model, each artificial neuron is substituted by a single qubit~\cite{Cao17}.
The neuron update is achieved with a quantum circuit that generates a non-linear activation function using control-flow feedback based on mid-circuit measurements.
This activation function is periodic, but locally resembles a sigmoid function.
Despite the mid-circuit measurement, this approach does not suffer from the collapse of relevant quantum information.
Rather, the measurement outcome signals that the neuron update is either successfully implemented or that a fixed, input independent operation is performed.
This other operation can be undone by feedback and the circuit rerun as necessary until success, leading to a constant, not exponential, overhead in the number of elementary operations required by RUS. Note that the overall fidelity of RUS circuits critically depends on the architecture and speed of the active feedback mechanism.

Our experiment uses 4 of the 7 transmons in a circuit QED processor~\cite{Versluis17} to implement a feedforward QNN with two inputs, one output and no intermediate layers.
We demonstrate that the QNN can learn each of the 16 2-to-1-bit Boolean functions by changing the weights and bias associated to the output neuron. 
It is particularly noteworthy that this architecture allows implementation of the XOR Boolean function using a single neuron, since this is a fundamental example of the limitations of classical artificial neuron constructions, which cannot capture the linear inseparability of such a function.

We follow the supervised learning paradigm, in which a set of training examples provides information to the network about the specific function to learn.
Our experiment uses multiple copies of a single input state (the maximal superposition of 4 inputs), demonstrating that the QNN can learn from a superposition.
Finally, we investigate the specificity of parameters learned for each of the Boolean functions by characterizing how well the values learned for one function can be used for any other.
This provides indications on using the QNN to discriminate between the Boolean functions when provided as a quantum black box.

\section{Results} \label{sec:results}

\textbf{Synthesizing non-linear functions using conditional gearbox circuits.}
The conditional gearbox circuit~\cite{Wiebe13} belongs to a class of RUS circuits~\cite{Paetznick14} that use one ancilla qubit $\QA$ and mid-circuit measurements to implement a desired operation.
The three-qubit version (Fig.~\ref{fig:conditional_gearbox}a) has input qubit $\QI$, output qubit $\QO$ and angles $w$ and $b$ as classical input parameters.
For an ideal processor starting with $\QA$ in $\kzA$, $\QI$ in computational state $\kkI$ ($k\in\{0,1\}$), and $\QO$ in arbitrary state $\kpO$, the coherent operations produce the state

\begin{equation}
\begin{gathered}
\kzA \kkI \kpO \rightarrow \sqrt{\psuccess(\thetak)} \kzA \kkI   R_x^{g(\thetak)} \kpO + \\
\sqrt{1-\psuccess(\thetak)} \koA \kkI   R_x^{-\frac{\pi}{2}} \kpO ,
\label{eq:gearbox_circuit}
\end{gathered}
\end{equation}
where $\thetak=kw+b$ and $\psuccess(\thetak)=\cos^4\left(\frac{\thetak}{2}\right)+\sin^4\left(\frac{\thetak}{2}\right)$.
A measurement of $\QA$ in its computational basis produces outcome $\mA=+1$ (projection to $\kzA$) with probability $\psuccess(\thetak)$. In this case,
the net effect on $\QO$ is a rotation around the $x$ axis of its Bloch sphere by angle $g(\thetak)$, where
\begin{equation}
g(\thetak)=2\arctan(\tan^2 \left(\frac{\thetak}{2} \right))
\label{eq:sigmoid_function}
\end{equation}
is a non-linear function with sigmoid shape (Fig.~\ref{fig:conditional_gearbox}d). This outcome constitutes success.

For failure (i.e., outcome $\mA=-1$ and projection onto $\koA$), the effect on $\QO$ is an $x$ rotation by $-\pi/2$, independent of $k$, $w$ and $b$.
In this case, the effect of the circuit can be undone using feedback, specifically $R_{x}^{\pi}$ and $R_{x}^{\frac{\pi}{2}}$ gates on $\QA$ and $\QO$, respectively.
The circuit can then be re-run with feedback corrections until success is achieved.
For an ideal processor, the average number of runs to success, $\mN$, is bounded by  $1\leq \mN=1/\psuccess(\thetak) \leq 2$.
This bound holds even when $\QI$ is initially in a superposition state $\kpI=\alpha\kzI+\beta \koI$.
In this general case, the output state upon success is still a superposition but with potentially different amplitudes:
\begin{equation}
\kzA \sum_{k=0}^{1} \alpha_k' \kkI R_{x}^{g(\thetak)}\kpO.
\end{equation}
The probability amplitudes can change, from $\alpha_k$ to $\alpha_k'$, depending on the initial $\kpI$, $w$, $b$, and $\NRTS$.
This distortion of probability amplitudes can be mitigated using amplitude amplification~\cite{Guerreschi19}, which we do not employ here.

\begin{figure}
\centering
\includegraphics[width=0.5\textwidth]{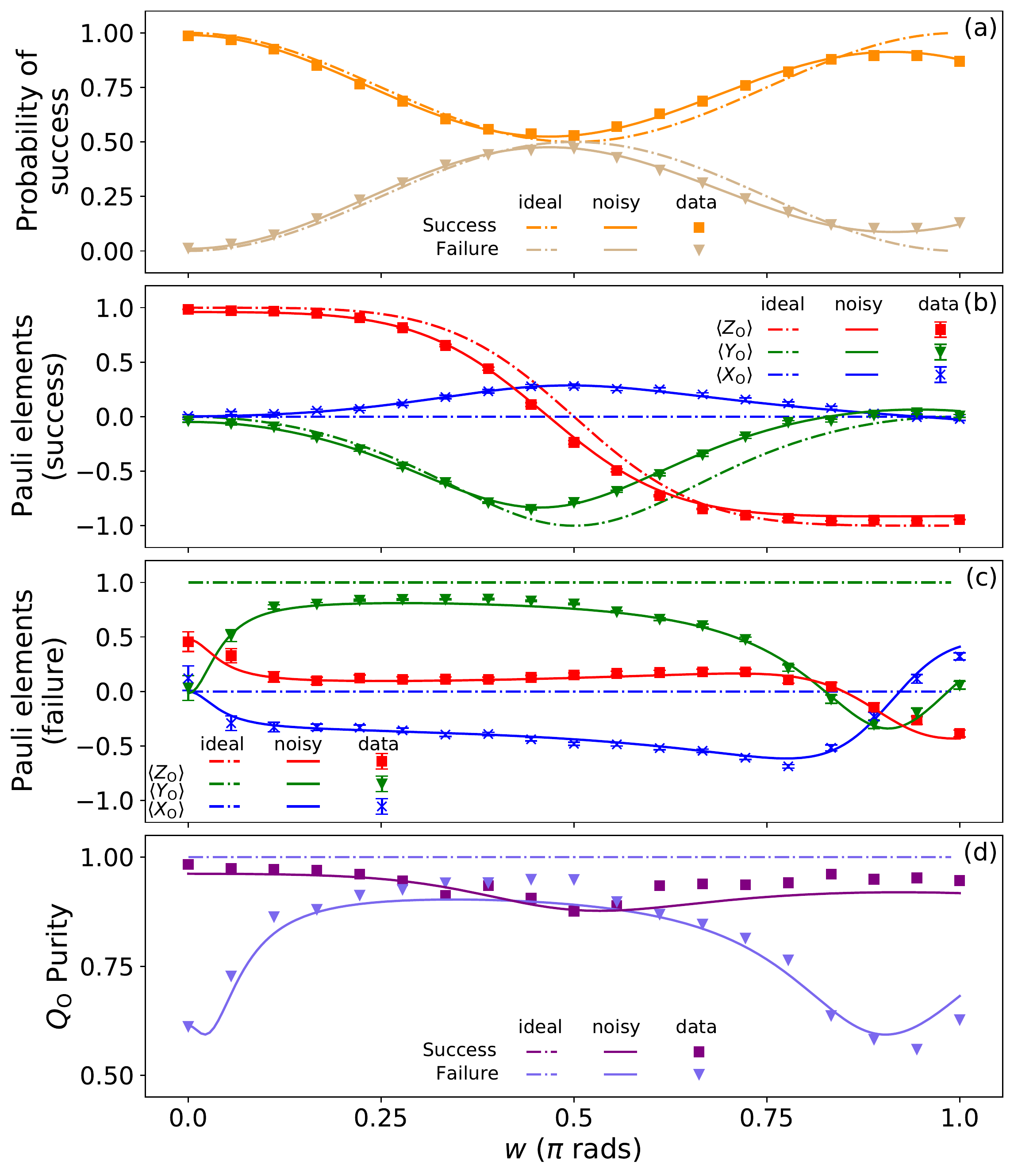}
\caption{
\textbf{Synthesis of non-linear functions using a conditional gearbox circuit.}
(a) Probability of success and failure at first iteration of the conditional gearbox circuit (Fig.~1) as a function of $w$ $(b=0)$.
(b,c) Pauli components of $\QO$ assessed by quantum state tomography conditioned on (b) success and (c) failure, for $\QI$ prepared in $\ket{1}$.
(d) Purity of $\QO$ for success and failure. All panels include experimental results (symbols), ideal simulation (dashed curves), and noisy simulation (solid curves).
}
\label{fig:sigmoid_synthesis}
\end{figure}

We compile the three-qubit conditional gearbox circuit into the native gate set of our processor (Fig.~\ref{fig:conditional_gearbox}b) and evidence its action after one round using state tomography of $\QO$ conditioned on success and failure. Figure~\ref{fig:sigmoid_synthesis} shows experimental results when preparing $\QI$ ($\QO$) in $\koI$ $(\kzO)$, setting $b=0$ and sweeping $w$, alongside simulation for both an ideal and a noisy processor. Qualitatively, the experimental results reproduce the key features of the ideal circuit: we observe a $\pi$-periodic oscillation in $\psuccess(w)$ with minimal value $0.5$ at $w=\pi/2$, and a sharp variation in $\ZO$ from $+1$ to $-1$ centered at $w=\pi/2$. However, the nonzero $\XO$ components observed for both success and failure indicate that the action on $\QO$ for both cases is not purely an $x$-axis rotation. The noisy simulation captures all key nonidealities observed. This simulation includes nonlinearity in single-qubit microwave driving, cross resonance~\cite{Chow11} effects between $\QA$ and $\QO$, phase errors in CZ gates, readout error in $\QA$, and qubit decoherence~\cite{SOM_QNN}.

\begin{figure*}
\centering
\includegraphics[width=\textwidth]{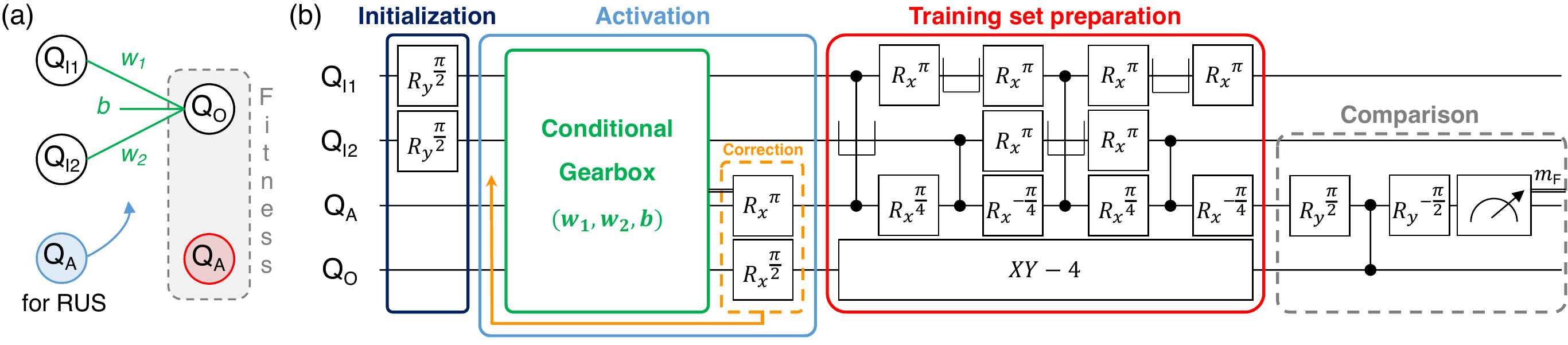}
\caption{
\textbf{Quantum neural network using the repeat-until-success conditional gearbox circuit.}
(a) Schematic representation of simplest feedforward network, highlighting the role played by parameters $(w_1, w_2, b)$ in weighing sum of input signals, before result is passed through non-linear activation function. $\QIone$ and $\QItwo$ are input nodes, $\QO$ is the output node and $\QA$ is an ancilla used first within the RUS circuit and then as expected output for the training set.
(b) Quantum circuit for a 3-neuron feedforward network.
This circuit is divided into four steps. Input ($\QIone$, $\QItwo$) preparation into maximal superposition; threshold activation into $\QO$ using RUS conditional gearbox circuit with $(w_1, w_2, b)$; unitary encoding of Boolean function (AND, in this case) using oracle; and comparison of $\QA$ with $\QO$. The symbol $\sqcup$ denotes parking of spectator qubit $\QItwo$ $[\QIone]$ during $\mathrm{CZ}(\QA,\QIone)$ $[\mathrm{CZ}(\QA,\QItwo)]$ gates~\cite{SOM_QNN}.
}
\label{fig:neural_network}
\end{figure*}

\begin{figure}
\centering
\includegraphics[width=0.5\textwidth]{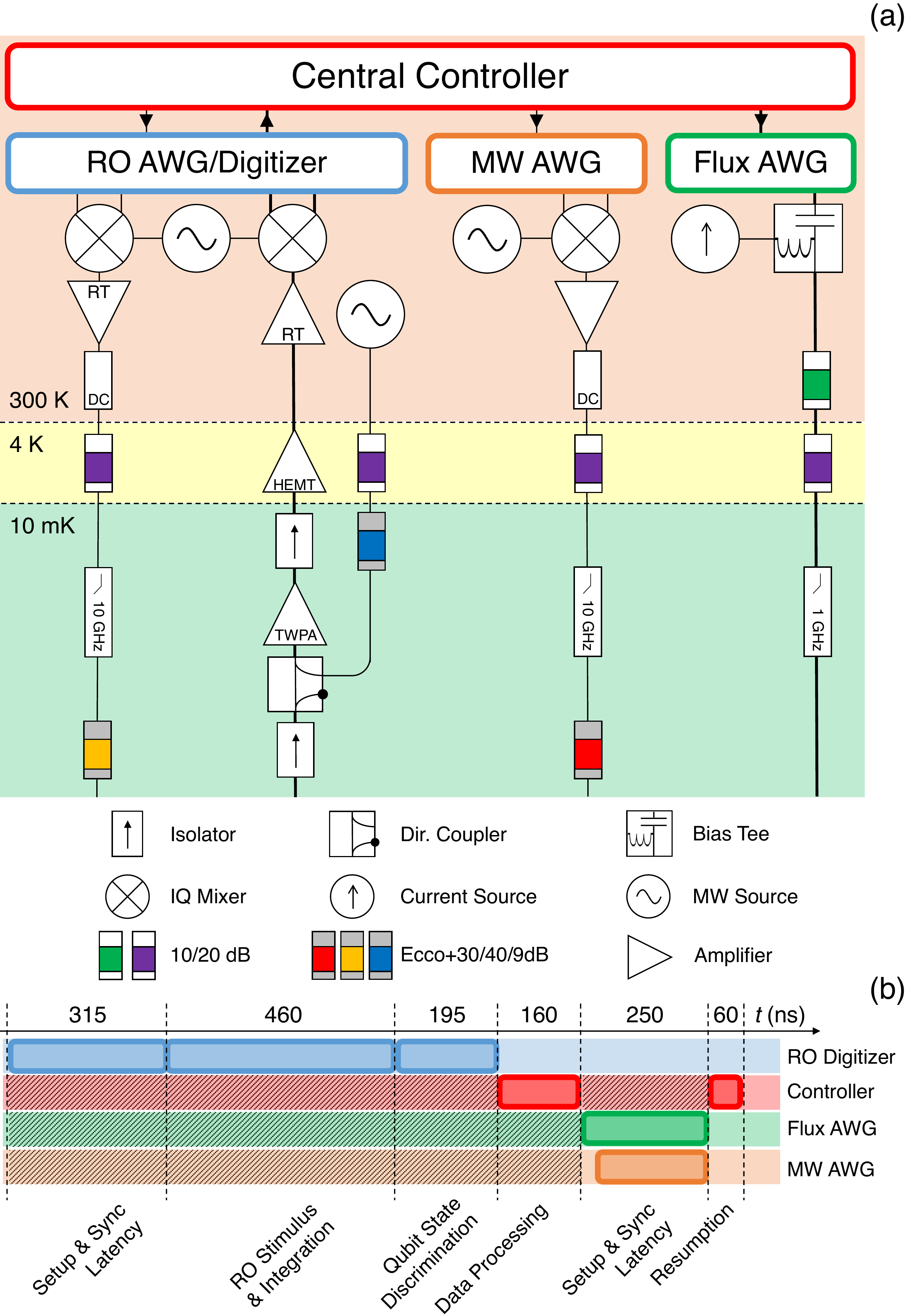}
\caption{\textbf{Quantum control setup.}
(a) Schematic of wiring and control electronics, highlighting critical feedback path between outputs of the quantum processor, the analog-interface devices, controller and the flux-drive lines;
(b) Timing diagram for the critical feedback path. Latency includes processing times necessary for synchronicity and hashed regions indicate idling operations for each instrument.}
\label{fig:control_setup}
\end{figure}

\textbf{Control-flow feedback on a programmable superconducting quantum processor.}
Active feedback is important for many quantum computing applications, including quantum error correction (QEC). Past demonstrations of QEC relied on the storage of measurements without real-time feedback~\cite{Krinner21,Chen21}. Moreover, real-time feedback has been demonstrated using data-flow mechanisms, where individual operations are applied conditionally~\cite{Andersen19}. In contrast, the implementation of RUS hinges on support for control-flow mechanisms in the control setup (Fig.~\ref{fig:control_setup}), where the entire sequence of operations has to be assessed and executed, depending on the results of measurements, in real-time.

In our quantum control architecture, a controller sequences the sets of operations to be performed in real-time, controlling various arbitrary waveform generators (AWG) and digitizers to implement the desired program. Therefore, our implementation of control-flow feedback focuses on this controller and achieves a maximum latency of 160~ns.
The latency to complete the full feedback loop of the overall control system (controller, analog-interface devices, and the entire analog chain) was measured to be 980~ns. This represents 3\% of the worst coherence time (see Table~\ref{tab:device_performance}), and sets an upper bound on the efficiency of RUS execution with the quantum processor. Further improvements could be achieved by optimizing the design of our RO AWG for trigger latency and speeding up the task of digital signal processing within the digitizers.

Note that the critical feedback path consists of the entire readout chain in addition to the slowest instrument, whose latency must also be accounted for before the branching condition is assessed and implemented. In our control setup, the slowest instrument is the Flux AWG, due to the latency introduced by various finite input response and exponential filters implemented in hardware for the correction of on-chip distortion of control pulses~\cite{Rol19}.

\textbf{Constructing a QNN using RUS circuits.}
The characteristic threshold shape of $g$ makes it useful in the context of neural networks: the conditional gearbox circuit can be seen as a non-linear activation function, whose rotations are controlled by the input qubits to mimic the propagation of information between network layers. We use these concepts~\cite{Cao17} to implement a minimal QNN capable of learning any of the 2-to-1-bit Boolean functions (see~\cite{SOM_QNN} for their definition and naming convention). These 16 functions can be separated into three categories (Table~\ref{tab:boolean_functions}): two constant functions (NULL and IDENTITY) have the same output for all inputs; 6 balanced functions (e.g., XOR) output 0 for exactly two inputs; and 8 unbalanced functions (e.g. AND) have the same output for exactly three inputs.

The 4-qubit circuit shown in Fig.~\ref{fig:neural_network} corresponds to a 3-neuron feedforward network.
Two quantum inputs ($\QIone$ and $\QItwo$) are initialized in a maximal superposition state.
Next, the RUS-based conditional gearbox circuit (now with three input angles $w_1$, $w_2$ and $b$) performs threshold activation of $\QO$. Following RUS (i.e., $\QA$ projected to $\kzA$), $\QA$ is reused for training set preparation. Here, the Boolean function $f$ is encoded in a quantum oracle mapping  $\kkIo \klIt \kzA \rightarrow \kkIo \klIt \ket{ f(k,l)_{\mathrm{A}}}$.
At this point, the 4-qubit register is ideally in state
\begin{equation}
\sum_{k,l=0}^1  \alpha_{kl}' \kkIo \klIt \ket{ f(k,l)_{\mathrm{A}}} R_x^{g(\thetakl)} \kzO,
\end{equation}
where $\thetakl = k w_1 + l w_2 + b$.
Finally, $\QA$ and $\QO$ are compared by mapping their parity onto $\QA$ and performing a final measurement on $\QA$ in the computational basis.

\begin{figure}[h!]
\centering
\includegraphics[width=0.5\textwidth]{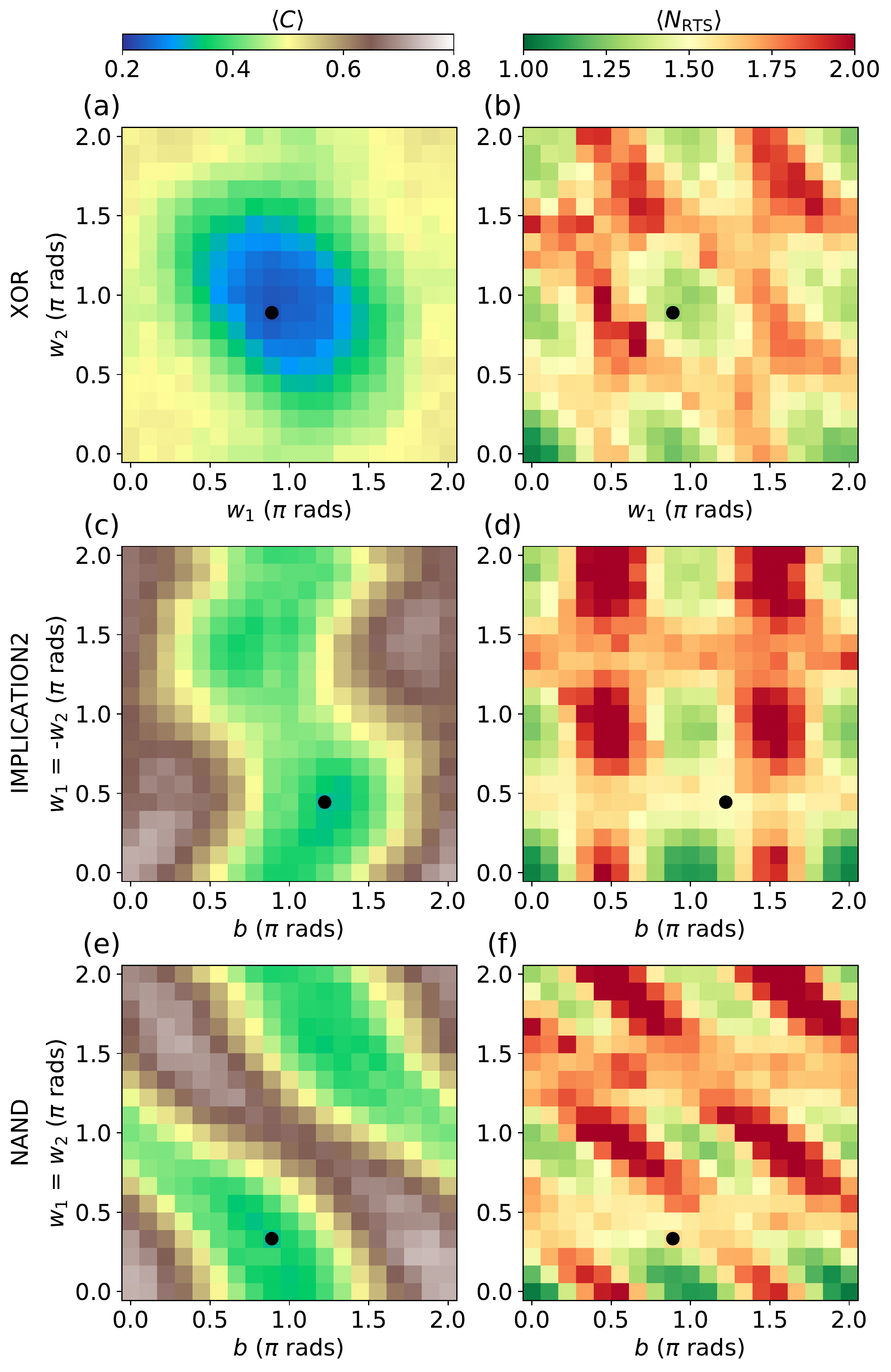}
\caption{
\textbf{Feature space landscapes of three Boolean functions.}
2-D slices of $\mC$  and $\mN$ for XOR (a, b), IMPLICATION 2 (c, d), and NAND (e, f). For each function, the slice
includes ($w_1$, $w_2$, $b$) parameters that minimize $\mC$ for an ideal quantum processor.
Black dots indicate the experimental parameters achieving minimal $\mC$ within each slice.
}
\label{fig:feature_spaces}
\end{figure}

We define $C=(1-\mA)/2$ from the output $\mA$ and estimate $\mC \in \left[0, 1\right]$ by averaging over 10,000 repetitions of the full circuit.
Training the QNN to learn a specific Boolean function thus amounts to minimizing $\mC$ over the 3-D input parameter space.
Beforehand, we explore the feature space landscapes. Figure~\ref{fig:feature_spaces} shows 2-D slices of $\mC$ and $\mN$ for three examples: XOR, IMPLICATION2 and NAND (see~\cite{SOM_QNN} for slices of all 16 functions).
These slices are chosen to include the optimal settings minimizing $\mC$ for an ideal quantum processor~\cite{SOM_QNN}.
These landscapes exemplify the complexity of the feature space and highlight the various symmetries and local minima that can potentially affect the efficient training of parameters.

\begin{figure}
\centering
\includegraphics[width=0.5\textwidth]{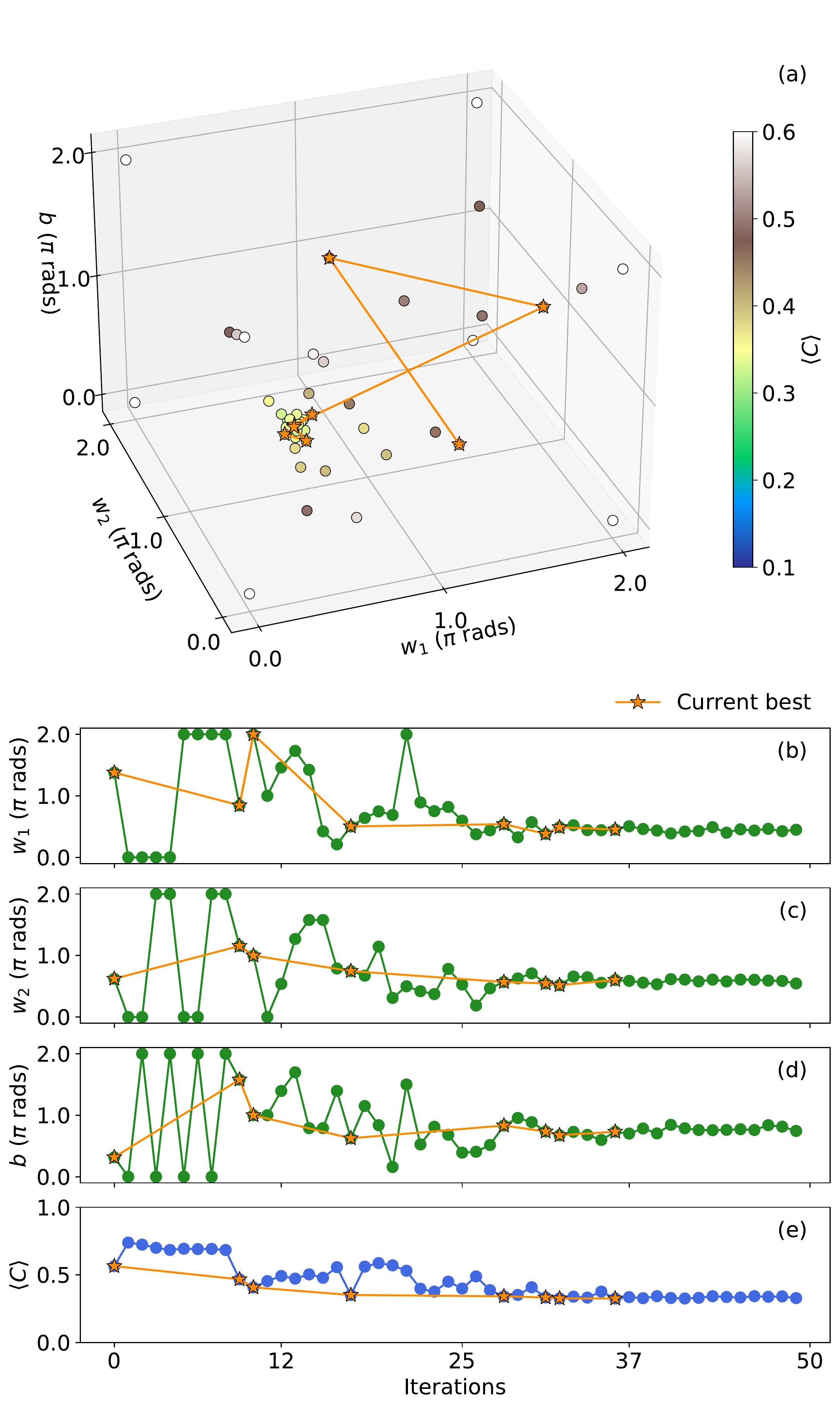}
\caption{\textbf{Learning the NAND function.}
(a) Training the QNN to learn NAND over the full parameter space $(w_1,w_2,b)$ by minimizing $\mC$ with an adaptive algorithm.
Training starts from a randomly-chosen point, then explores the boundaries, and ultimately converges within $\sim50$ steps.
(b-e) Evolution of training parameters $(w_1,w_2,b)$ and $\mC$  as a function of training step. The current best setting achieved is marked by a star.
}
\label{fig:training_nand}
\end{figure}

\textbf{Training a QNN from superpositions of data.}
To train the QNN, we employ an adaptive learning algorithm~\cite{Nijholt19} to minimize $\mC$ over the full 3-D parameter space.
Figure~\ref{fig:training_nand} shows the training process for NAND, chosen for the complexity of its feature space.
The parameters evolve with each training step, starting from a randomly chosen initial point, then exploring the bounds, and subsequently converging to the global minimum in $\sim50$ training steps.
This satisfactory behavior is observed for all the Boolean functions.

\begin{figure}
\includegraphics[width=0.5\textwidth]{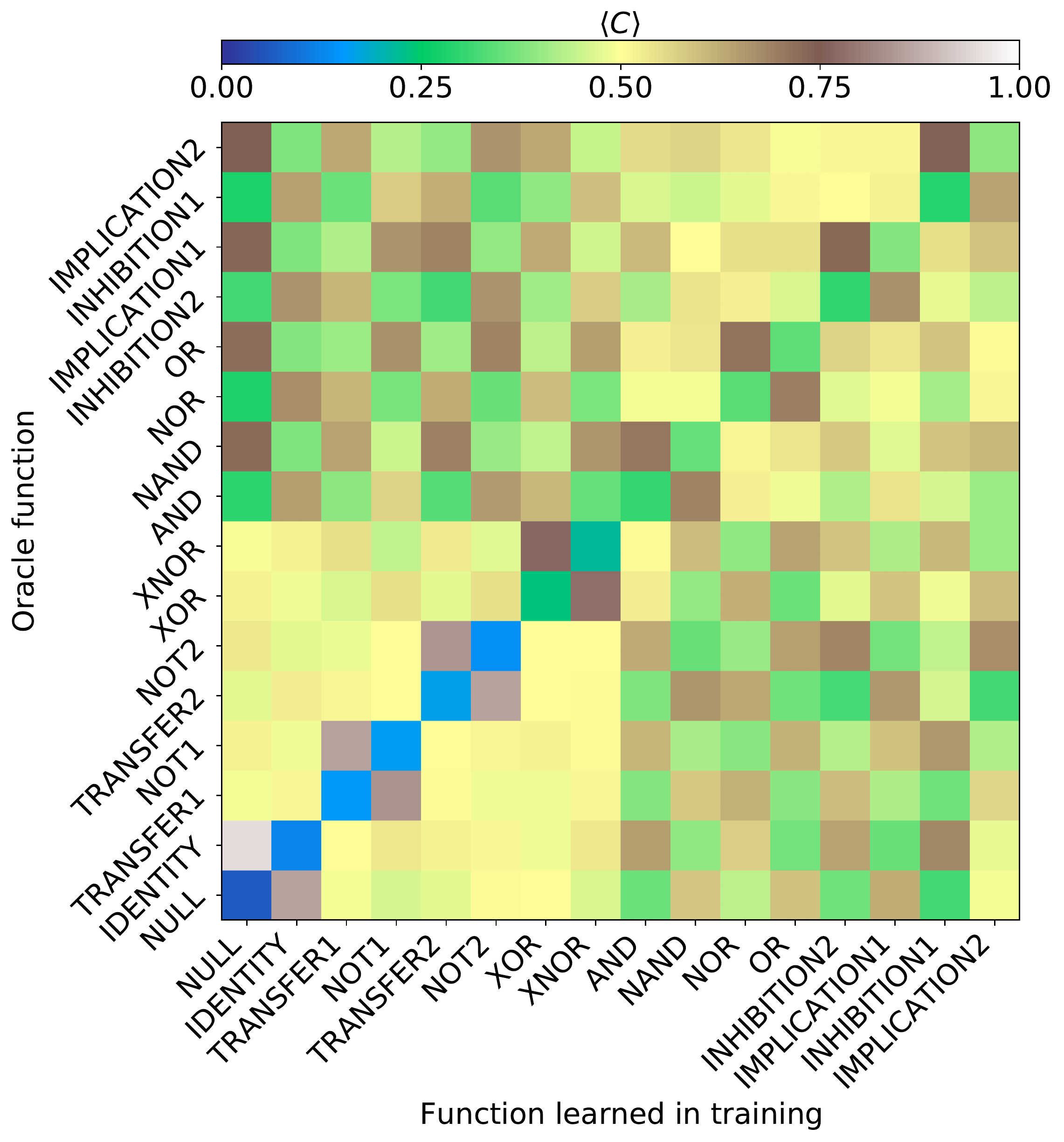}
\caption{\textbf{Specificity of the quantum neural network.}
Cost function of the optimized parameter set for every training function (horizontal axis) against all oracle functions (vertical axis).
In each axis, the functions are ordered from constant, to balanced, to unbalanced. Functions are put alongside their complementary function (NULL and IDENTITY, TRANSFER1 and NOT1, etc.).
For an ideal processor, $\mC$ values are expected at or close to multiples of 0.25, due to the varying overlap between the 16 Boolean functions (i.e., the number of 2-bit inputs producing different 1-bit outcome).
Further differences arise in experiment due to variation in the average circuit depth of the RUS-based activation functions and in the fixed circuit depth of oracle functions.
}
\label{fig:specificity_matrix}
\end{figure}

Following training of the QNN for each Boolean function, we investigate the specificity of learned parameters by preparing the 256 pairs of trained parameters and function oracles and measuring $\mC$ for each pair.
To understand the structure of the experimental specificity matrix (Fig.~\ref{fig:specificity_matrix}), it is worthwhile to first consider the case of an ideal processor (see~\cite{SOM_QNN}).
Along the diagonal, we expect $\mC=0$ for constant and balanced functions, which can be perfectly learned, and $\mC\approx0.029$ for unbalanced functions, which cannot be perfectly learned due to the finite width of the activation function $g(\theta)$. For off-diagonal terms, we expect $\mC$ at or close to multiples of 0.25, the multiple being set by the number of 2-bit inputs for which the paired training function and oracle function have different 1-bit output. For example,  NAND and XOR have different output only for input $00$, while TRANSFER1 and NOT1, which are complementary functions, have different output for all inputs. Note that every constant or balanced function, when compared to any unbalanced function, has different output for exactly two inputs. Evidently, while the described pattern is discerned in the experimental specificity matrix,  deviations result from the compounding of decoherence, gate-calibration, crosstalk, and measurement errors. These errors affect the 256 pairs differently for two main reasons. First, the average circuit depth of the RUS-based conditional gearbox circuit is higher for unbalanced functions. Second, the fixed circuit depth of oracles is also significantly higher for unbalanced functions, as these all require a Toffoli-like gate which we realize using CZ and single-qubit gates. Noisy simulation~\cite{SOM_QNN} modeling the main known sources of error in our processor produces a close match to Fig.~\ref{fig:specificity_matrix}.

Despite the evident imperfections, we have shown that it is possible to train the network across all functions, arriving at parameters that individually optimize each landscape. The circuit is thus able to learn different functions using multiple copies of a single training state corresponding to the superposition of all inputs, despite the complexity of feature space landscapes for various Boolean functions.

\section{Discussion} \label{sec:discussion}

We have seen that RUS is an effective strategy to address the probabilistic nature of the conditional gearbox circuit, allowing the deterministic synthesis of non-linear rotations. Even at the error rates of current superconducting quantum processors, it allowed the implementation of a QNN that reproduced a variety of classical neural network mechanisms while preserving quantum coherence and entanglement. Moreover, we have shown that this QNN architecture could be trained to learn all 2-to-1-bit Boolean functions using superpositions of training data.

This minimal QNN represents a fundamental building block that can be used to build larger QNNs. With larger numbers of qubits, these neurons could form  multi-layer feed-forward networks containing hidden layers between inputs and outputs. Beyond feedforward networks, this minimal QNN is amenable to the implementation of various other network architectures, from Hopfield networks to quantum autoencoders~\cite{Cao17}.

Finally, this work highlights the importance of real-time feedback control performed within the qubit coherence time and the quantum-classical interactions governing RUS algorithms. The ability to implement RUS circuits is in itself a useful result, as the active feedback architecture demonstrated is crucial for various other applications of a quantum computer, including active-reset protocols and the synthesis of circuits of shorter depth relative to purely unitary circuit design~\cite{Paetznick14}, of value in areas such as quantum chemistry. Moreover, recent work into quantum error correction (QEC) highlights the importance of real-time quantum control in protocols for the distillation of magic states or, when coupled to a real-time decoder, the correction of errors. Similarly to real-time feedback, the construction of a real-time decoder that meets the stringent requirements for QEC with superconducting qubits requires application-specific hardware developments that are the focus of ongoing work.

\section*{Author contributions}
	M.S.M. performed the experiment and data analysis.
	M.B., N.H. and L.D.C. designed the device.
	N.M., C.Z. and A.B. fabricated the device.
	M.S.M., J.F.M. and H.A. calibrated the device.
	M.S.M., W.V., J.S., J.S. and C.G.A. designed the control electronics.
	G.G.G. and L.D.C performed the numerical simulations and motivated the project.
	M.S.M., G.G.G. and L.D.C. wrote the manuscript with input from A.Y.M., S.P.P. and X.Z.
	A.Y.M. and L.D.C. supervised the theory and experimental components of the project, respectively.

\section*{Acknowledgements} \label{sec:acknowledgements}
    We thank G.~Calusine and W.~Oliver for providing the traveling-wave parametric amplifiers used in the readout amplification chain. This research is supported by Intel Corporation and by the Office of the Director of National Intelligence (ODNI), Intelligence Advanced Research Projects Activity (IARPA), via the U.S. Army Research Office Grant No. W911NF-16-1-0071. The views and conclusions contained herein are those of the authors and should not be interpreted as necessarily representing the official policies or endorsements, either expressed or implied, of the ODNI, IARPA, or the U.S. Government.

\section*{Data Availability}
	The data shown in all figures of the main text and supplementary material are available at \url{http://github.com/DiCarloLab-Delft/Quantum_Neural_Networks_Data}. 

\section*{Competing Interests}
	The authors declare no competing interests.


\end{bibunit}

\onecolumngrid
\clearpage
\renewcommand{\theequation}{S\arabic{equation}}
\renewcommand{\thefigure}{S\arabic{figure}}
\renewcommand{\thetable}{S\arabic{table}}
\renewcommand{\bibnumfmt}[1]{[S#1]}
\renewcommand{\citenumfont}[1]{S#1}
\setcounter{figure}{0}
\setcounter{equation}{0}
\setcounter{table}{0}
\begin{bibunit}[naturemag]
\section*{Supplemental material for 'Realization of a quantum neural network using repeat-until-success circuits in a superconducting quantum processor'}

This supplement provides additional information in support of statements and claims made in the main text.

\setcounter{section}{0}
\section{Device characteristics}
\label{si:sec:device-characteristics}

\begin{figure}[!h]
\centering
\includegraphics[width=0.45\textwidth]{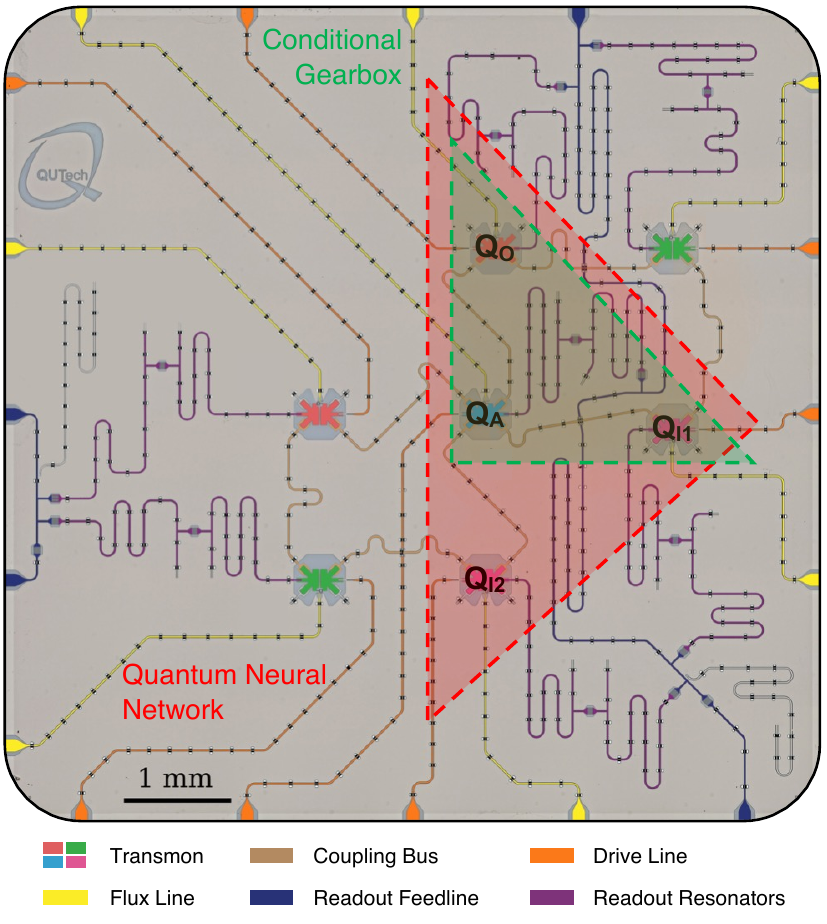}
\caption{
\textbf{Superconducting quantum processor with overlaid circuit topology.}
Optical image of the quantum processor with added falsecolor to emphasize different circuit QED elements. Qubit names are also overlaid to indicate the four transmons used in this work.
The green (red) patch shows the transmons used in the 3-qubit conditional gearbox circuit (Figs.~\ref{fig:conditional_gearbox} and~\ref{fig:sigmoid_synthesis}) and in the QNN (Figs.~\ref{fig:neural_network} to ~\ref{fig:specificity_matrix}).
}
\label{fig:device_topology}
\end{figure}

The device used is already introduced and described in prior published experiments~\cite{Sagastizabal20,Negirneac20,Marques22}.
Select metrics for the four transmon qubits used in this work are provided in Table.~\ref{tab:device_performance}.
Figure ~\ref{fig:device_topology} highlights the circuit QED elements allowing coherent control and measurement.
Each qubit has a dedicated flux-control line, microwave drive line, and readout resonator with dedicated Purcell filter.
Readouts of the four qubits employed in this experiment use a single common feedline.
We note that $\QA$ is driven from this feedline due to an issue with its dedicated microwave drive line.
This leads to cross-resonance effects during single-qubit gates of $\QA$.
The extra amplification required to overcome the filtering effect of the readout and Purcell resonators also leads to non-linearity when driving $\QA$ (Section~\ref{si:sec:simulation-methodology}).

\begin{table}[!h]
\begin{tabular}{ccccc}
\hline
Qubit               & $\QO$ & $\QIone$ & $\QItwo$ & $\QA$ \\
\hline
Qubit transition frequency at sweetspot, $\omega_q/2\pi$ ($\GHz$) & 6.433 & 4.534 & 4.562 & 5.887 \\
Transmon anharmonicity, $\alpha/2\pi$ ($\MHz$)                    & -270  & -314  & -312  & -294  \\
Readout frequency, $\omega_r/2\pi$ ($\GHz$)                       & 7.492 & 6.913 & 6.646 & 7.058 \\
Relaxation time, $\Tone$ ($\us$)                                  & 34    & 39    & 82    & 67    \\
Ramsey dephasing time, $\Ttwostar$ ($\us$)                        & 41    & 16    & 60    & 63    \\
Echo dephasing time, $\Ttwoecho$ ($\us$)                          & 53    & 84    & 106   & 72    \\
Multiplexed readout fidelity, $F_{\mathrm{RO}}$ (\%)              & 99.2  & 99.9  & 99.5  & 98.9  \\
Residual excitation, $r$ (\%)                                     & 0.0  & 3.1  & 4.7  & 0.6  \\
Single-qubit gate fidelity, $F_{\mathrm{1Q}}$ (\%)                & 99.95 & 99.91 & 99.97 & 99.90 \\
CZ gate fidelity, $F_{\mathrm{2Q}}$ (\%)                          & 99.7  & 97.5  & 97.0  & ---   \\
CZ gate Leakage, $\leakrate$ (\%)                                 & 0.6  & 0.8  & 0.5  & ---   \\
\hline
\end{tabular}
\caption{
\textbf{Summary of select parameters and performance metrics of the four transmon qubits used in the experiment.}
Coherence times are obtained using standard time-domain measurements~\cite{Krantz19}. 
The multiplexed readout fidelity, $F_{\mathrm{RO}}$, is the average assignment fidelity extracted from single-shot readout histograms~\cite{Bultink18} .
The single-qubit gate fidelity, $F_{\mathrm{1Q}}$, is extracted from individual single-qubit randomized benchmarking.
The two-qubit gate fidelity, $F_{\mathrm{2Q}}$, is obtained through interleaved randomized benchmarking with modifications to quantify leakage, $\leakrate$~\cite{Magesan12,Wood18}.
With the exception of frequency values, quantities listed are vulnerable to drift. 
For example, relaxation and dephasing times typically vary by several $\us$ and readout fidelity and residual excitation vary by a few percent. 
}
\label{tab:device_performance}
\end{table}

\begin{figure}[!h]
\centering
\includegraphics[width=0.4\textwidth]{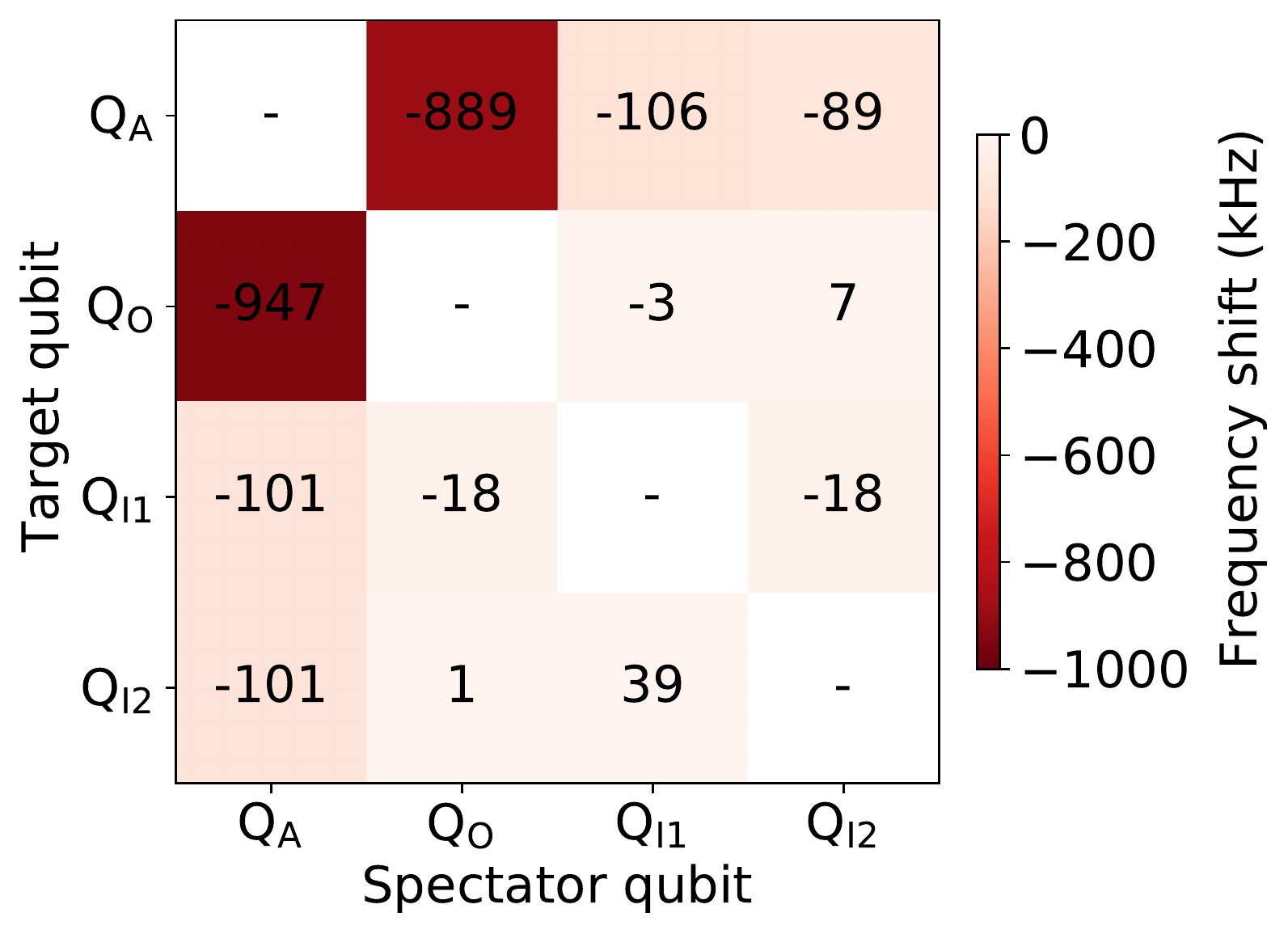}
\caption{\textbf{Residual $ZZ$ coupling.}
Characterization of residual $ZZ$ coupling between all qubit pairs at the bias point (simultaneous flux sweetspot). The matrix elements indicate the shift in frequency experienced by one qubit (target qubit) when another (spectator qubit) changes from $\ket{0}$ to $\ket{1}$. The procedure used for this measurement is similar to the one described in~\cite{Sagastizabal20}.
}
\label{fig:residual_zz}
\end{figure}

\section{2-to-1-bit Boolean functions}
\label{si:sec:boolean-functions}

\begin{table}[!h]
\begin{tabular}{c|c|c|c|c|c|c}
\hline
\multirow{2}{*}{Name}    & \multirow{2}{*}{Definition} & \multicolumn{4}{c|}{Truth table}              & \multirow{2}{*}{Characteristic}  \\
                         &                             & $0_1 0_2$ & $0_1 1_2$ & $1_1 0_2$ & $1_1 1_2$ &                                  \\
\hline
NULL               & 0                                 &0&0&0&0        & Constant\\
IDENTITY           & 1                                 &1&1&1&1        & Constant\\
TRANSFER 1         & $I_{1}$                           &0&0&1&1        & Balanced\\
NOT 1              & $\overline{I_{1}}$                &1&1&0&0        & Balanced\\
TRANSFER 2         & $I_{2}$                           &0&1&0&1        & Balanced\\
NOT 2              & $\overline{I_{2}}$                &1&0&1&0        & Balanced\\
XOR                & $I_{1}\oplus I_{2}$               &0&1&1&0        & Balanced\\
XNOR               & $\overline{I_{1} \oplus I_{2}}$   &1&0&0&1        & Balanced\\
AND                & $I_{1}\wedge I_{2}$               &0&0&0&1        & Unbalanced\\
NAND               & $\overline{I_{1} \wedge I_{2}} $  &1&1&1&0        & Unbalanced\\
NOR                & $\overline{I_{1} \vee I_{2}}$     &1&0&0&0        & Unbalanced\\
OR                 & $I_{1}\vee I_{2}$                 &0&1&1&1        & Unbalanced\\
INHIBITION 2       & $ \overline{I_{1}} \wedge I_{2}$  &0&1&0&0        & Unbalanced\\
IMPLICATION 1      & $I_{1}\vee \overline{I_{2}}$      &1&0&1&1        & Unbalanced\\
INHIBITION 1       & $I_{1}  \wedge \overline{I_{2}}$  &0&0&1&0        & Unbalanced\\
IMPLICATION 2      & $\overline{I_{1}} \vee  I_{2}$    &1&1&0&1        & Unbalanced\\
\hline
\end{tabular}

\caption{
\textbf{2-to-1-bit Boolean functions.}
Naming, definition, truth table, and characteristic of the 16 2-to-1-bit Boolean functions.
The functions have bit inputs $I_{1}$ and $I_{2}$.
Symbols $\overline{\ }$, $\oplus$, $\wedge$, and $\vee$ denote NOT, XOR (exclusive or), AND, and OR operations, respectively.
Constant functions have the same output for all inputs.
Balanced functions output 0 for exactly two inputs.
Unbalanced functions have the same output for exactly three inputs.
}
\label{tab:boolean_functions}
\end{table}

\begin{figure}[!h]
    \centering
    \includegraphics[width=1\textwidth]{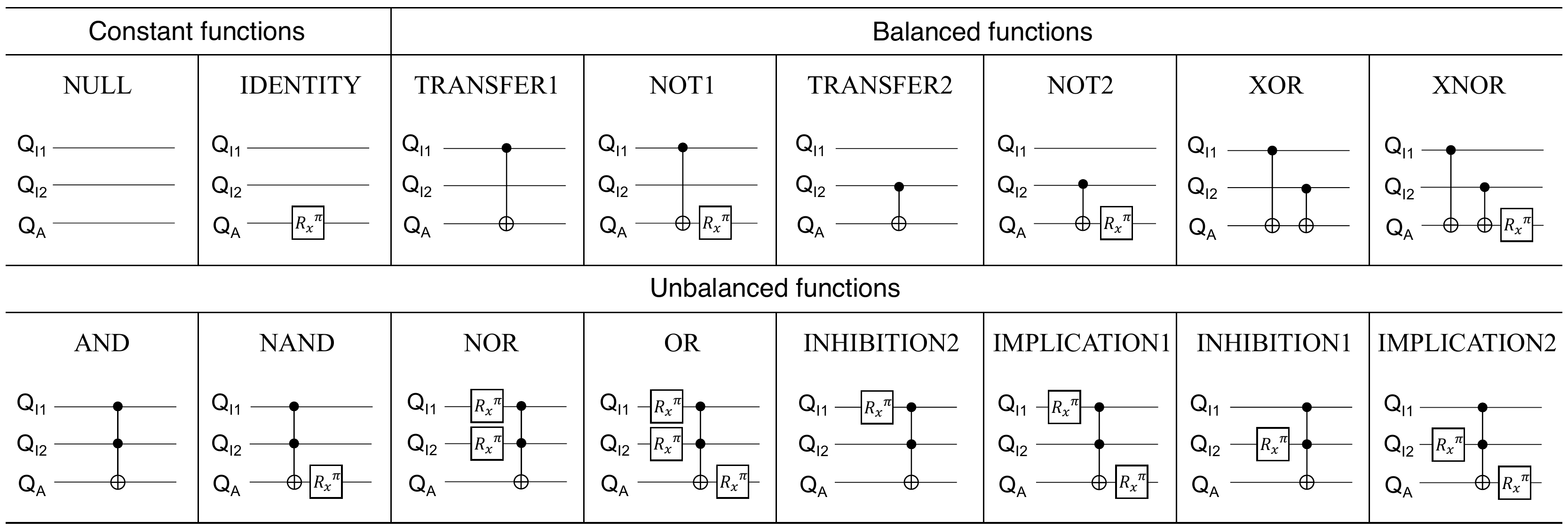}
    \caption{\textbf{Training set preparation circuits for 2-to-1-bit Boolean functions.}
            Three-qubit circuits (inputs $\QIone$ and $\QItwo$, and output $\QA$) implementing oracles for the preparation of training sets of each 2-to-1-bit Boolean function. The circuits here are not written in the native gate set. When compiling them into the native gate set, we perform additional simplifications and add dynamical decoupling pulses for error mitigation.}
    \label{fig:boolean_oracles}
\end{figure}

The definition and nomenclature used for the 16 2-to-1-bit Boolean functions are presented in Table~\ref{tab:boolean_functions}. The corresponding quantum oracles needed for the preparation of training datasets are presented in Fig.~\ref{fig:boolean_oracles}. These circuits are compiled using the native gate set of the processor, making simplifications wherever possible. For example, we substitute all CC-NOT gates with CC-$i$X gates (Fig.~\ref{fig:toffoli_decomposition}) as they can be implemented with lower circuit depth. This is possible as $\QIone$ and $\QItwo$ are not reused after training set preparation in the QNN circuit (Fig.~\ref{fig:neural_network}) and, therefore, the difference between CC-NOT and CC-$i$X gates is not relevant in this context.

\section{Error mitigation strategies}
\label{si:sec:error-mitigation}

\subsection{Characterization and optimization of CZ gates}

\begin{figure}[!h]
    \centering
    \includegraphics[width=0.75\textwidth]{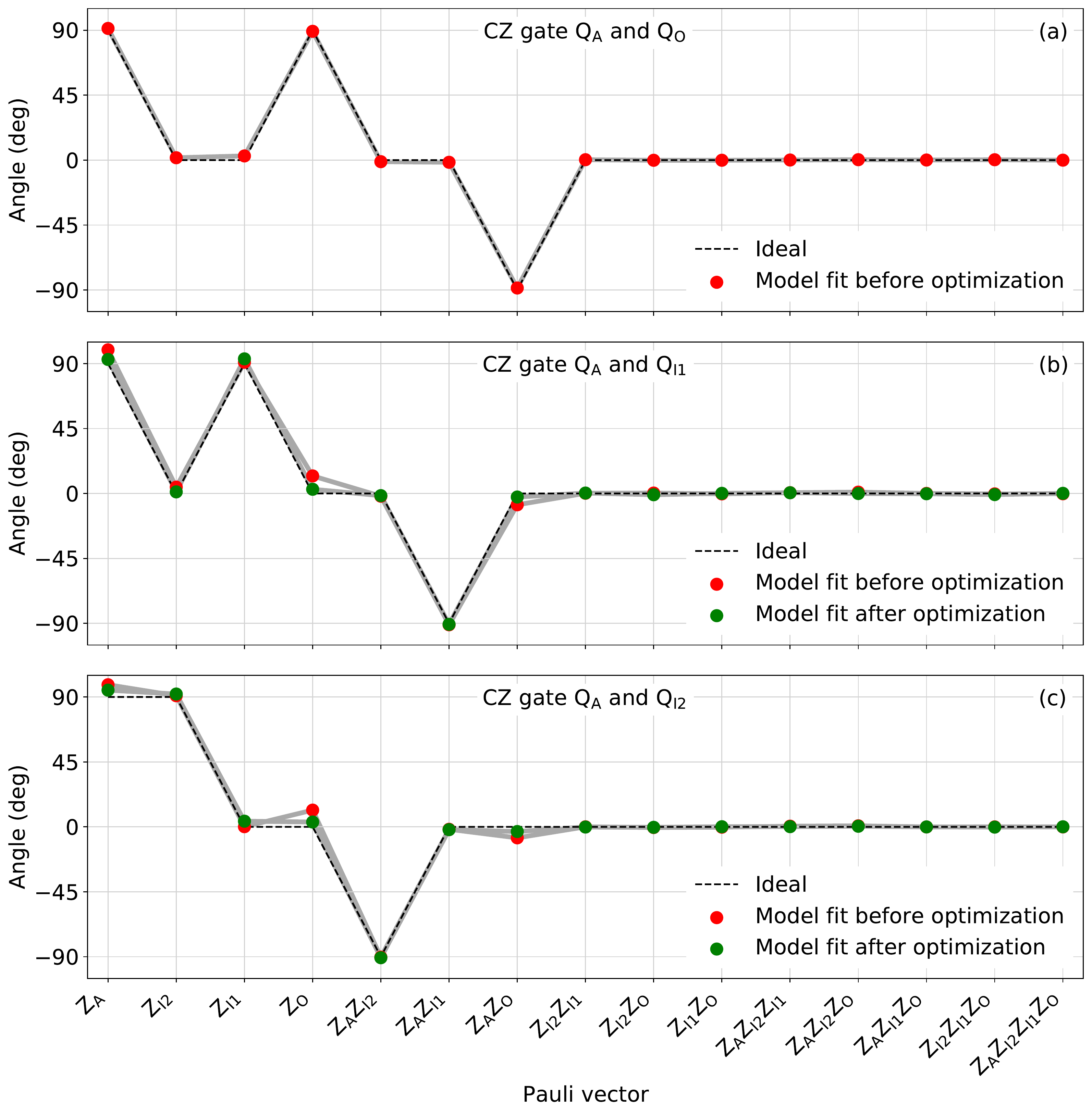}
    \caption{\textbf{Characterization of native two-qubit gates.}
            Characterization of the phase action of the CZ gates between $\QA$ and each of $\QO$, $\QIone$, and $\QItwo$, separated into one-, two-, three- and four-qubit phase terms.}
    \label{fig:characterization_cz}
\end{figure}

As observed in previous work ~\cite{Sagastizabal20,Negirneac20,Marques22} using this quantum processor, the residual $ZZ$ coupling between qubit pairs constitute a significant source of error. This translates to spectator qubits coupling to either of the qubits involved in a CZ gate, leading to the increase of leakage and phase errors when spectators are not in $\ket{0}$. To assess the phase impact of spectators, we fit the action of each CZ gate (between pairs $\QA$-$\QO$, $\QA$-$\QIone$ and $\QA$-$\QItwo$ to the model
\begin{equation}
\begin{gathered}
U = e^{-\frac{i}{2}\theta_{\mathrm{A}}\ZA} e^{-\frac{i}{2}\theta_{\mathrm{I2}}\ZIt} e^{-\frac{i}{2}\theta_{\mathrm{I1}}\ZIo} e^{-\frac{i}{2}\theta_{\mathrm{O}}\ZO} \\
\times e^{-\frac{i}{2}\theta_{\mathrm{A, I2}}\ZA\ZIt} e^{-\frac{i}{2}\theta_{\mathrm{A, I1}}\ZA\ZIo} e^{-\frac{i}{2}\theta_{\mathrm{A, O}}\ZA\ZO} e^{-\frac{i}{2}\theta_{\mathrm{I2, I1}}\ZIt\ZIo} e^{-\frac{i}{2}\theta_{\mathrm{I2, O}}\ZIt\ZO} e^{-\frac{i}{2}\theta_{I1, O}\ZIo\ZO} \\
\times e^{-\frac{i}{2}\theta_{\mathrm{A, I2, I1}}\ZA\ZIt\ZIo} e^{-\frac{i}{2}\theta_{\mathrm{A, I2, O}}\ZA\ZIt\ZO} e^{-\frac{i}{2}\theta_{\mathrm{A, I1, O}}\ZA\ZIo\ZO} e^{-\frac{i}{2}\theta_{\mathrm{I2, I1, O}}\ZIt\ZIo\ZO} e^{-\frac{i}{2}\theta_{\mathrm{A, I2, I1, O}}\ZA\ZIt\ZIo\ZO}.
\label{eq:cz-model}
\end{gathered}
\end{equation}
This model includes all single-, two-, three-, and four-qubit phase terms. To extract the 15 terms, we first measure the quantum phase imparted on each qubit for each of the 8 computational states of the other three qubits and then perform a least-squares fit to the model.  Results are shown in
Fig.~\ref{fig:characterization_cz}. We observe single-qubit and two-qubit phase errors for $\mathrm{CZ}(\QA,\QIone)$ and $\mathrm{CZ}(\QA,\QItwo)$, and particularly on terms $\ZO$, $\ZA$ and $\ZO\ZA$. 
These are consistent with measurements of residual $ZZ$ couplings between all qubit pairs in this quantum processor (Fig.~\ref{fig:residual_zz}), which show strongest coupling between $\QA$ and $\QO$.

To mitigate these phase errors, two $R_x^{\pi}$ gates are performed on $\QO$ back-to-back during $\mathrm{CZ}(\QA,\QIone)$ and $\mathrm{CZ}(\QA,\QItwo)$ (Fig.~\ref{fig:conditional_gearbox}b). 
This is done to symmetrize the population of the spectator qubits during the CZ gates while having the added gates compile to identity, leaving the overall effect of the circuit unchanged.
The addition notably reduces phase errors (Fig.~\ref{fig:characterization_cz}b-c).
Characterization of $\mathrm{CZ}(\QA,\QO)$  (Fig.~\ref{fig:characterization_cz}a) showed accurate performance without similar error mitigation, which is likely due to low residual couplings of both $\QItwo$ and $\QIone$ to both $\QA$ and $\QO$. 

We note that these phase errors are time varying and were captured here after CZ-gate calibration. Simulation efforts described below extracted different values for these angles, hinting at drift between calibration and data collection for the experiments.

\subsection{Characterization and compensation of drive non-linearity in $\QA$}

\begin{figure}[!h]
    \centering
    \includegraphics[width=0.5\textwidth]{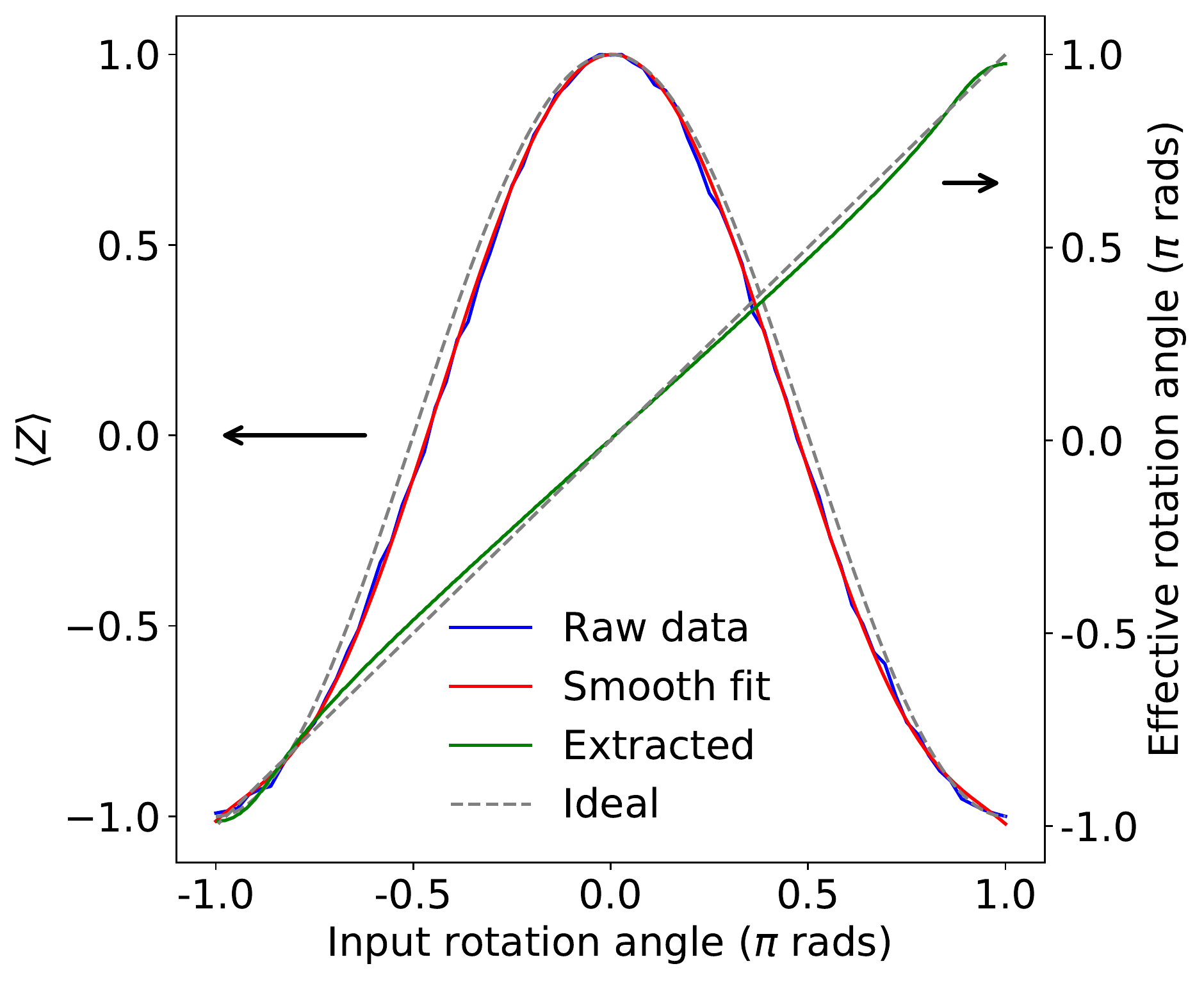}
    \caption{\textbf{Characterization of arbitrary-angle rotations of $\QA$.}
            Characterization of microwave control on $\QA$ through the effective rotation angle implemented on the qubit. Small inaccuracies stemming from non-linearity of the microwave-drive chain can be accounted for in this way, to ensure fine control is implemented. These measurements are carried out while preparing all spectator qubits in $\ket{0}$.
            }
    \label{fig:microwave_characterization}
\end{figure}

To ensure proper calibration of the arbitrary single-qubit rotations required, despite known non-linearities associated with the amplifiers and microwave-drive lines required for the implementation of these gates, the Rabi oscillation of $\QA$ is thoroughly characterized using quantum state tomography (Fig.~\ref{fig:microwave_characterization}). Using this dataset, the effective rotation angle of $\QA$ is computed and used to correct for these effects. Despite our best efforts, errors consistent with over-rotations on $\QA$ are still observed in the horizontal compression evident in Fig.~\ref{fig:sigmoid_synthesis}).

\subsection{Characterization and optimization of CC-iX gate circuit}

\begin{figure}[!h]
    \centering
    \includegraphics[width=0.45\textwidth]{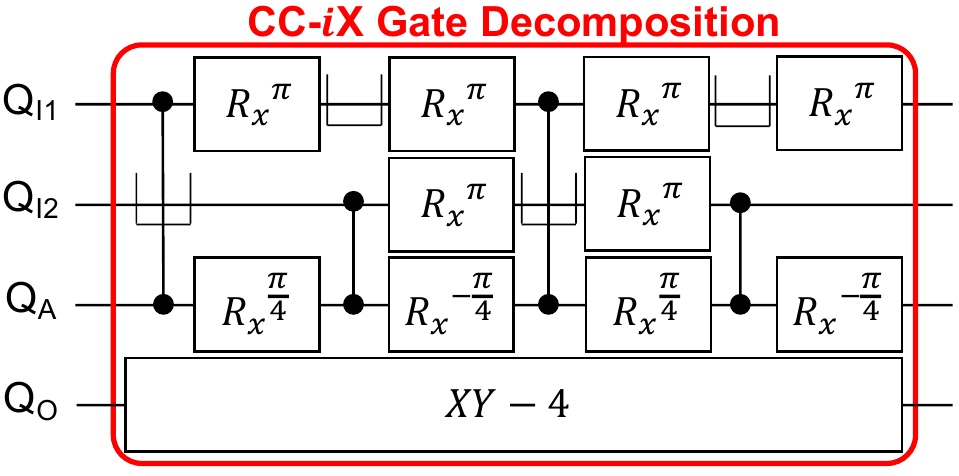}
    \caption{
    \textbf{CC-$i$X gate decomposition.}
    Decomposition of the CC-$i$X gate into the native gate set of the quantum processor. To maximize fidelity, dynamical decoupling pulses are added to mitigate the effect of residual ZZ coupling.
    }
    \label{fig:toffoli_decomposition}
\end{figure}

The implementation of oracles for unbalanced Boolean functions requires three-qubit  operations (Fig.~\ref{fig:boolean_oracles}). We can use the CC-$i$X gate (Fig.~\ref{fig:toffoli_decomposition}) as a proxy to the CC-NOT (Toffoli) gate, which can be implemented with lower depth. The difference between CC-$i$X and CC-NOT is only a two-qubit phase that is of no relevance in the QNN circuit (Fig.~\ref{fig:neural_network}).

\begin{figure}[!h]
    \centering
    \includegraphics[width=\textwidth]{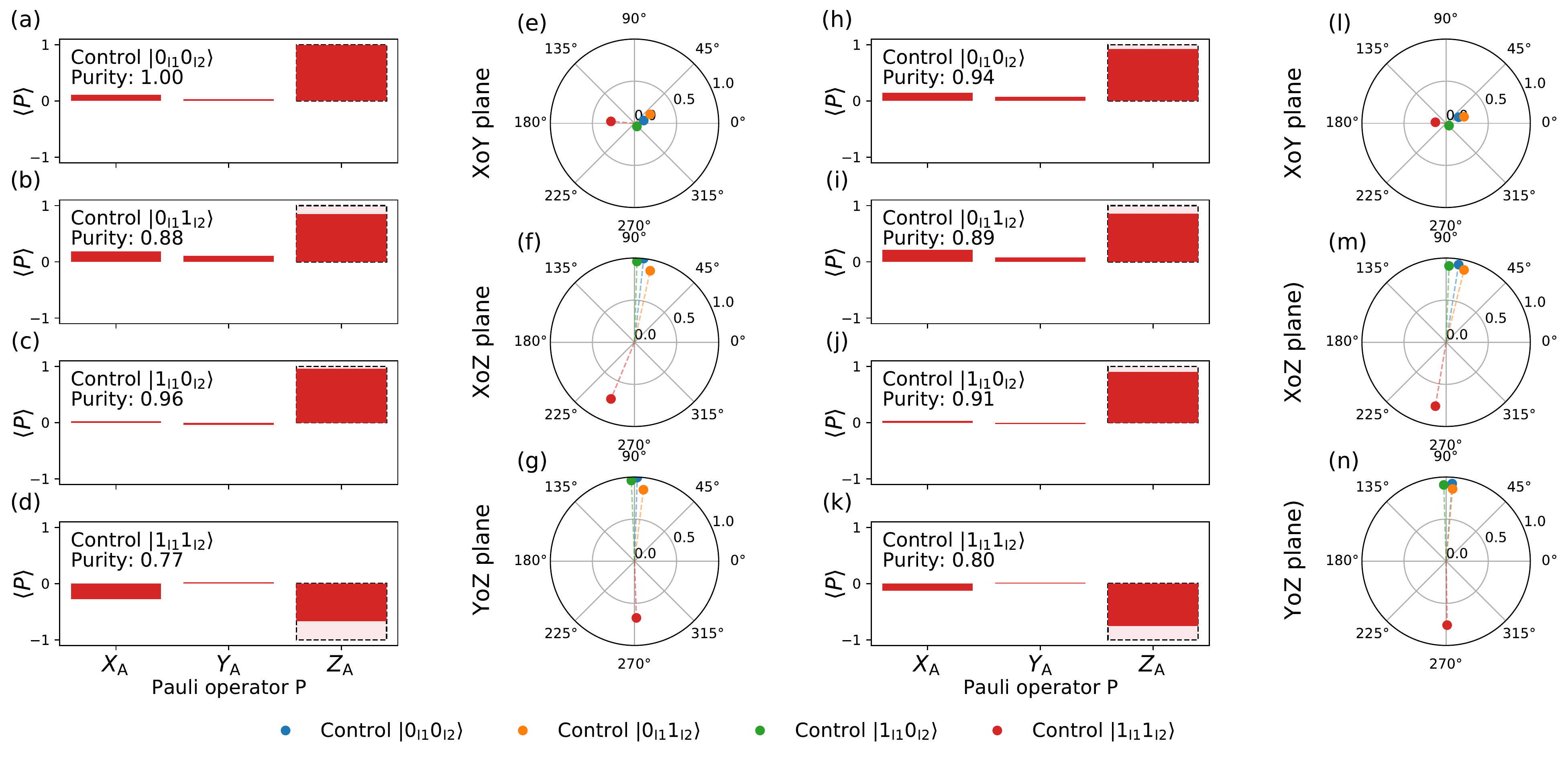}
    \caption{\textbf{Characterization of CC-$i$X gate decomposition.}
            Full tomography of $\QA$ after circuit implementing CC-$i$X gate before (a-g) and after (h-n) optimization meant to symmetrize the population of $\QIone$ and $\QItwo$ during the circuit.
            }
    \label{fig:toffoli_characterization}
\end{figure}

The effect of this circuit on $\QO$ is characterized through tomography for various input states (Fig.~\ref{fig:toffoli_characterization}a). In particular, the result of optimizing the circuit against residual $ZZ$ effects by symmetrizing the population of $\QIone$ and $\QItwo$ using $R_x^{\pi}$ gates is studied (Fig.~\ref{fig:toffoli_characterization}b). This optimization produced only minor improvements, most likely owing to the reduced residual $ZZ$ couplings observed between $\QA$, $\QIone$ and $\QItwo$.

\subsection{Characterization of RUS correction pulse}

\begin{figure}[!h]
    \centering
    \includegraphics[width=0.5\textwidth]{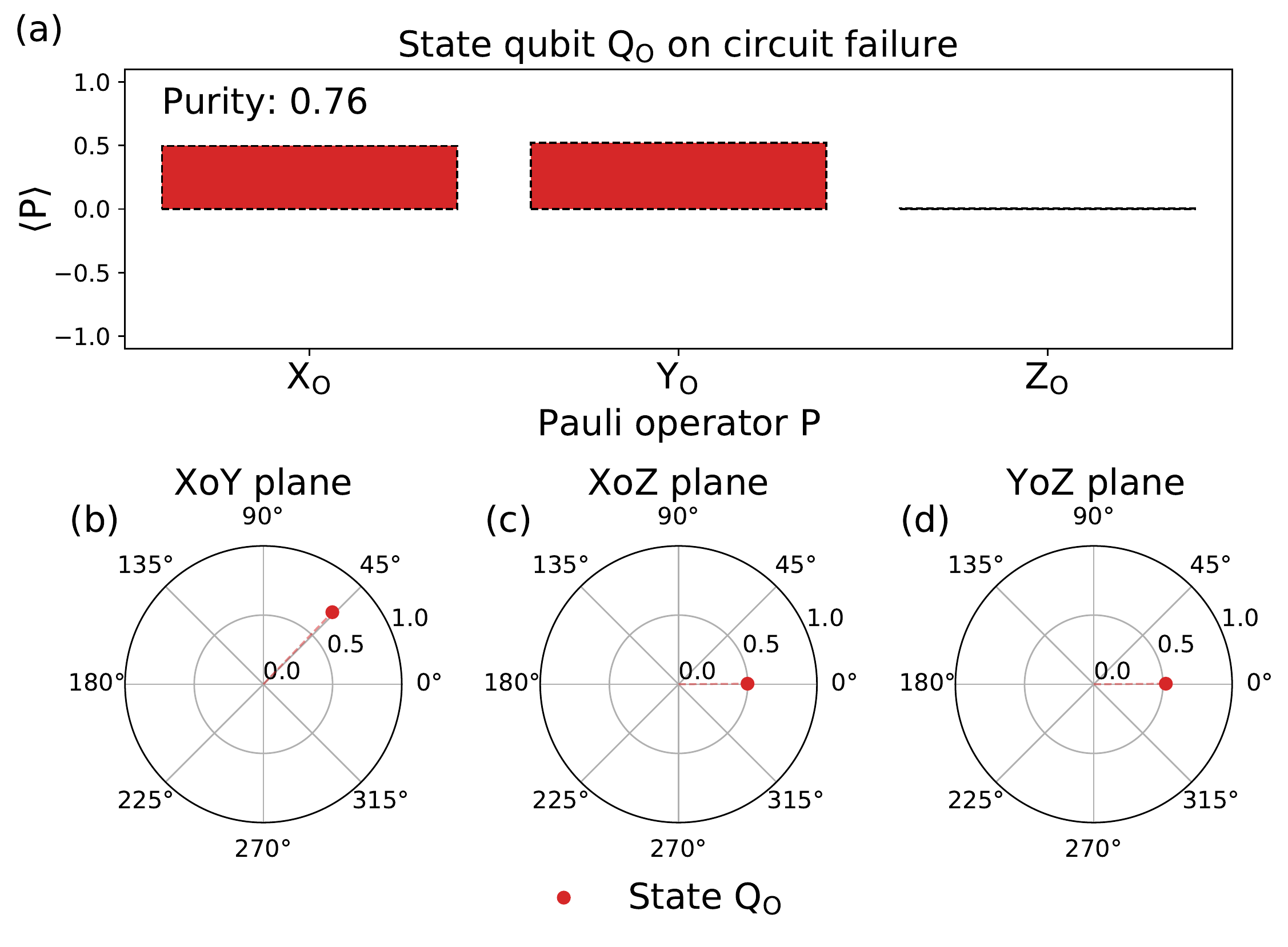}
    \caption{\textbf{Characterization of correction pulse.}
            Characterization of $\QO$ after one iteration of the RUS conditional gearbox circuit, performed through full tomography conditioned on failure. This experiment allowed the calibration of an optimal pulse to recover $\QO$, before another iteration of the conditional gearbox circuit is attempted. This characterization highlights a coherent phase error in $\QO$.
            }
    \label{fig:correction_pulse}
\end{figure}

The use of the gearbox circuit with RUS is contingent on the ability to recover $\QA$ and $\QO$ in case of failure. This can be done with  $R_{x}^{\pi}$ and $R_{x}^{\frac{\pi}{2}}$ rotations, respectively. However, inaccuracies in the CZ gates stemming from residual $ZZ$ couplings lead to a dependency of the optimal $\QO$ correction pulse on $w_1$ and $w_2$. This is studied further with recourse to simulation using realistic parameters extracted from hardware (Fig.~\ref{fig:sigmoid_synthesis}). Furthermore, spectator toggling effects during the idling time of $\QO$ are expected to lead to a coherent rotation of the qubit, effectively changing the axis of the $\frac{\pi}{2}$ rotation required to bring the qubit to $\ket{0}$.

To characterize the optimal correction pulse, tomography is performed on $\QO$ after running the conditional gearbox circuit through the first measurement (Fig.~\ref{eq:gearbox_circuit}b) and post-selecting  on $m=-1$. To maximize the probability of failure, therefore increasing the significance of the results acquired, this measurement is performed for $(w_1, w_2, b) = (\pi, -\pi, \pi/2)$. The results (Fig.~\ref{fig:correction_pulse}) showed a coherent effective rotation $R_{-40\degrees}^{-\frac{\pi}{2}}$, therefore leading the correction pulse to be defined as $R_{-40\degrees}^{\frac{\pi}{2}}$.

\clearpage
\section{Cost function of networks for all 2-to-1-bit Boolean functions}
\label{si:sec:network-fitness}

\begin{figure}[!h]
\centering
\includegraphics[width=0.69\textwidth]{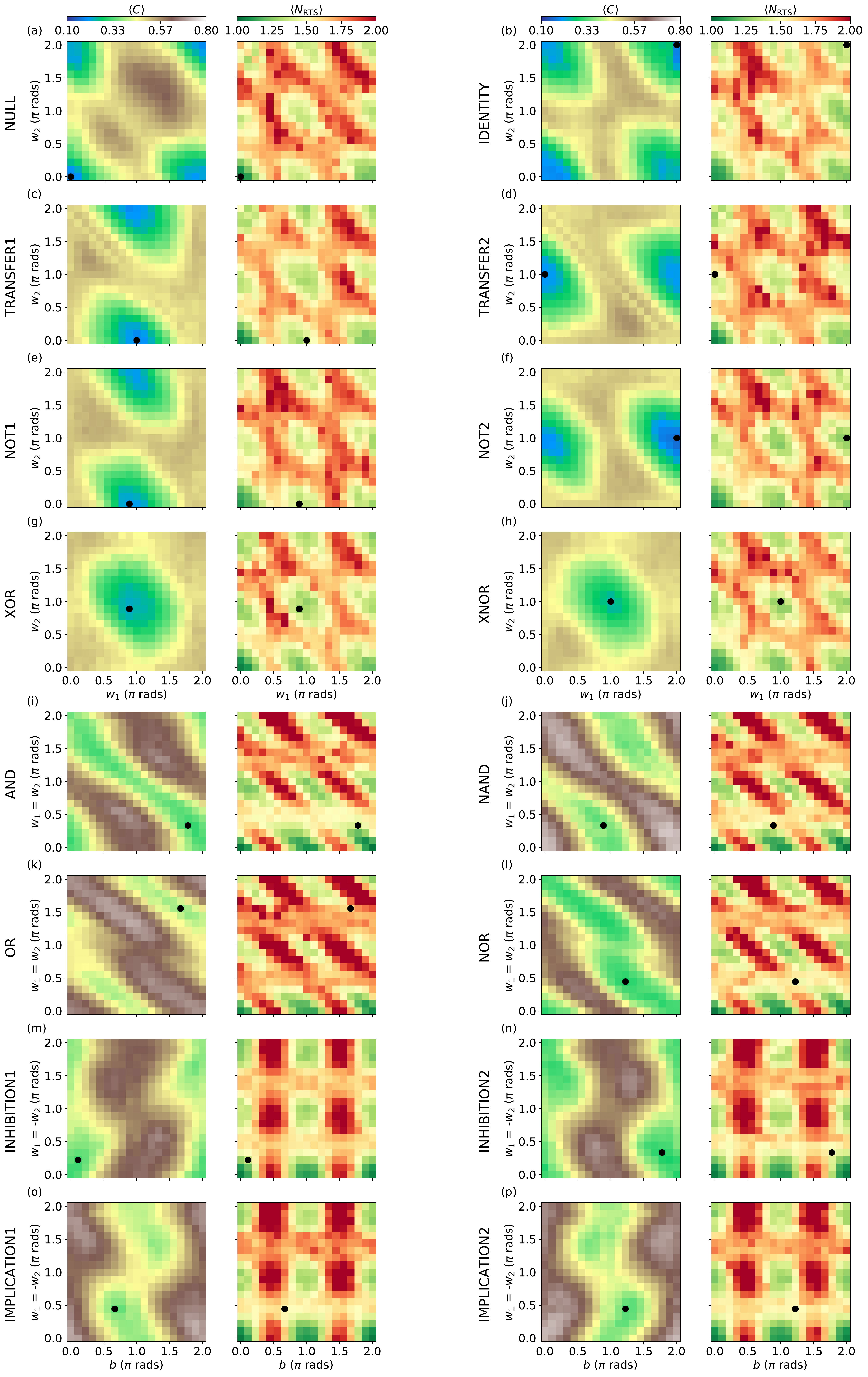}
\caption{\textbf{Experimental feature space landscapes of all 2-to-1-bit Boolean functions.}
2-D slices of $\mC$ (left panels) and $\mN$ (right panels) for all Boolean functions.
For each function, the slice is chosen to include the optimal parameters $(w_1, w_2, b)$ that minimize $\mC$ for an ideal quantum processor.
Black dots indicate the experimental parameters achieving minimal $\mC$ within each slice.
}
\label{fig:all_landscapes}
\end{figure}

\begin{figure}
    \centering
    \includegraphics[width=0.5\textwidth]{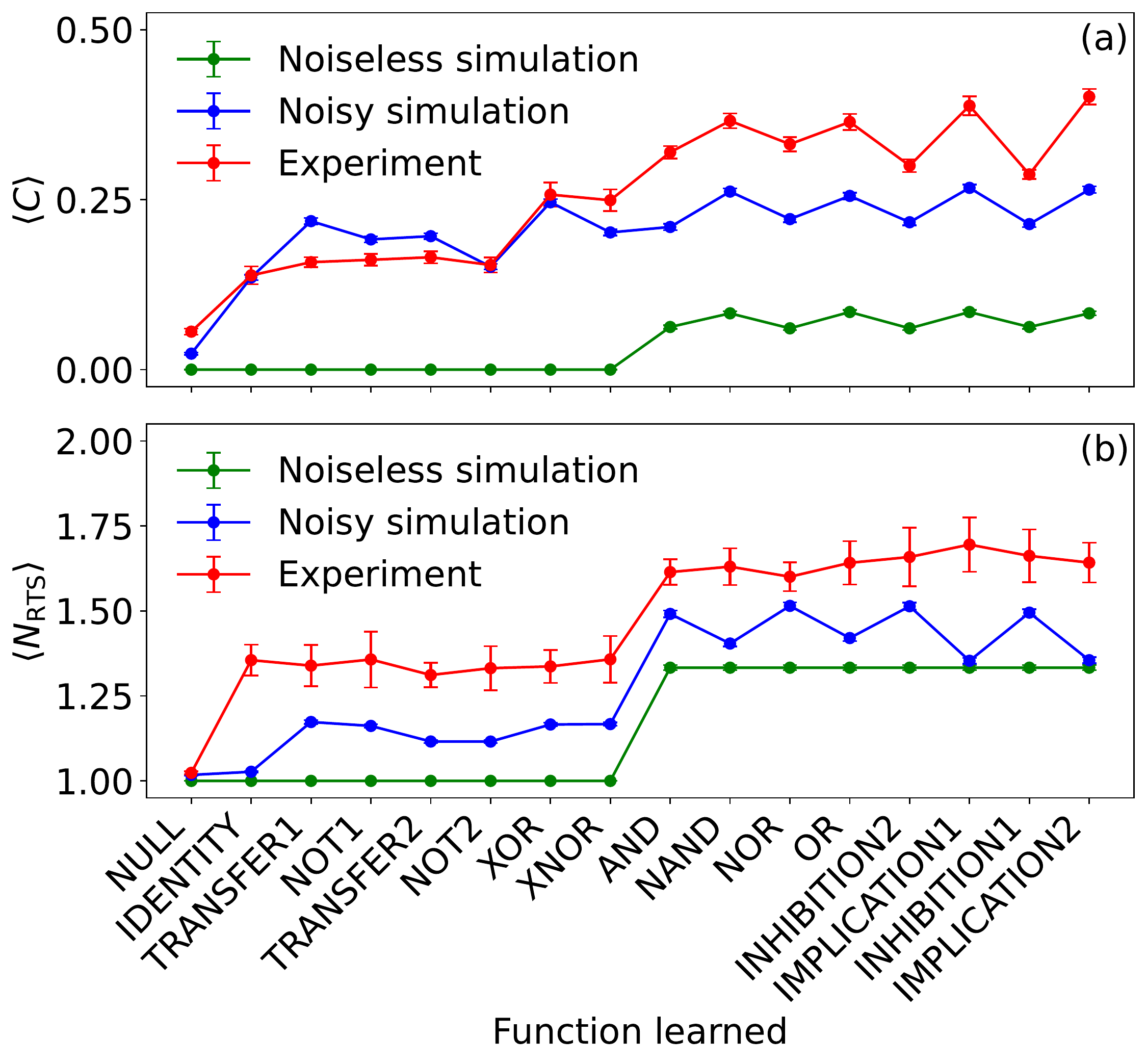}
    \caption{\textbf{Metrics of quantum-neural-network training.}
             (a)$\mC$  and (b) $\mN$ with optimized parameters for each Boolean function. Error bars represent the standard deviation over 50 function evaluations, each based on 8000 repetitions. These data correspond to the diagonal of the specificity matrix (Fig.~\ref{fig:specificity_matrix}). The increased $\mC$ observed for unbalanced functions (right half) results from higher circuit depth in both the implementation of the function oracles as well as the RUS-based conditional gearbox circuit (note the higher $\mN$ for these functions).
            }
    \label{fig:statistics_specificity}
\end{figure}

\pagebreak
\section{Comparing gearbox zero iteration, one iteration and original activation}
\label{si:sec:comparison-ghearbox}

\begin{figure}[!h]
    \centering
    \includegraphics[width=0.75\textwidth]{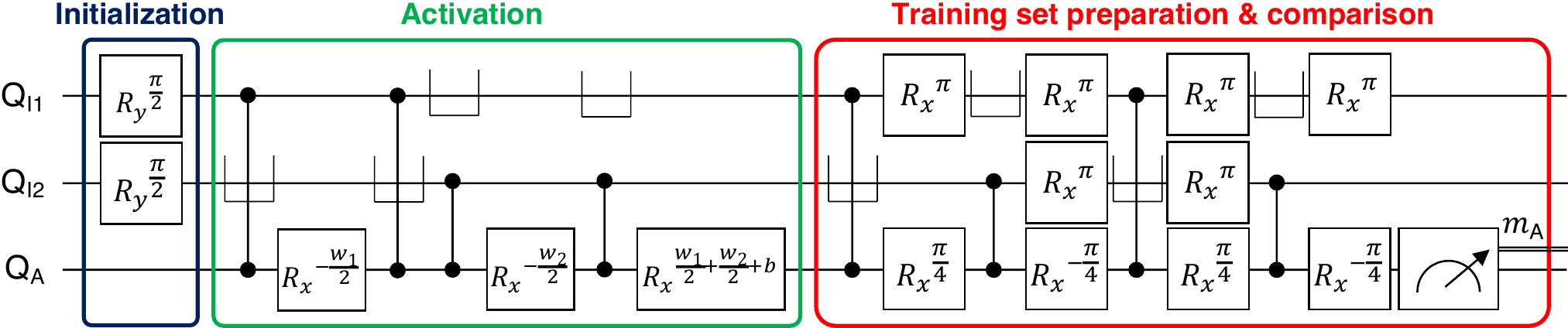}
    \caption{\textbf{Quantum neural network with sinusoidal activation.}
        3-qubit circuit implementing a Rabi activation function with control parameters $(w_1, w_2, b)$. The circuit takes advantage of the Rabi oscillation to implement a softer activation function through controlled $R_{x}^{w_1}$, $R_{x}^{w_2}$ and $R_{x}^{b}$ gates on $\QA$. This circuit is deterministic and, contrary to the circuit presented in Fig.~\ref{fig:conditional_gearbox}, does not require an ancilla qubit. Instead, the training set preparation is effected directly on $\QA$, after which $\mC$ is assessed through $m_A$.
        }
    \label{fig:rabi_activation}
\end{figure}

\begin{figure}[!h]
    \centering
    \includegraphics[width=0.5\textwidth]{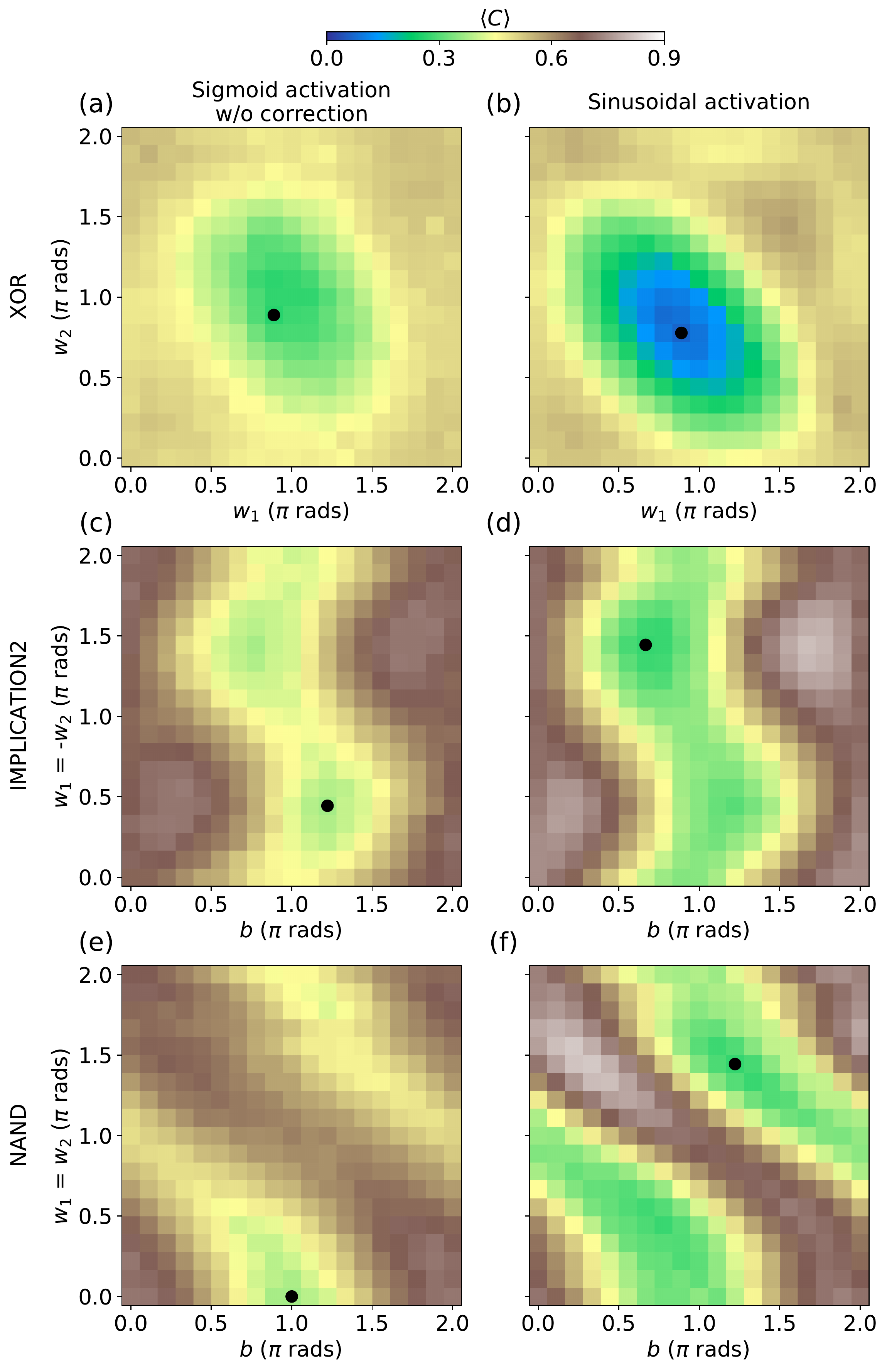}
    \caption{\textbf{Comparison of feature space landscapes obtained for variations of the activation function circuits.}
        Landscapes of $\mC$ for functions XOR (a, b), IMPLICATION 2 (c, d), and NAND (e, f), representative of different Boolean functions, obtained with variations of the activation function circuit. (a, c, e) represent activation through a single iteration of the gearbox circuit (sigmoid-like activation), without correction on failure, and (b, d, f) use the Rabi oscillation (sinusoidal-like activation) as a softer non-linear function. These represent 2-D slices of the 3-D landscapes, chosen at specific cuts where simulation indicated the minimum to be located. The points corresponding to minimal $\mC$ for each of these cuts are represented in all subplots.
        }
    \label{fig:activation_landscapes}
\end{figure}

To study the effectiveness of correction and the practical usefulness of employing a RUS strategy with the gearbox  circuit, a variation of the gearbox circuit is implemented such that $\QA$ always completes after the first iteration, with no correction in case of failure. Furthermore, following the observation that a simple Rabi oscillation has non-linear $\mZ(\theta)$ (although $g(\theta)=\theta$ for an analogy to Fig.~\ref{fig:conditional_gearbox}), we propose a 3-qubit version of the quantum neuron that is capable of using this property as its activation function. Although such a circuit follows a slightly modified construction (Fig.~\ref{fig:rabi_activation}), it should still be able to implement a three-neuron feedforward network with parameters $(w_1, w_2, b)$, without needing an extra ancillary qubit. However, this should come at the cost of a softer activation function. Indeed, the difference between sinusoidal and sigmoid-like activation functions gives the original QNN circuit an advantage in ideal simulation.

The effectiveness of the two new circuit variations in experiment (no-correction and Rabi activation) is illustrated through their feature space landscapes (Fig.~\ref{fig:activation_landscapes}) for XOR, IMPLICATION 2 and NAND, a set representative of the complexity of feature spaces for all the functions considered. For functions whose minimum is expected with parameters for which ideally $\mN=1$ (XOR is the only such example here), all circuit variations appear to work equally well. However, for functions whose minimum is expected for parameters leading to a higher $\mN$, the no-correction circuit variation already highlights several distortions, leading to minima that privilege always outputting one regardless of its inputs, i.e., $w_1=w_2=0$ and $b=\pi$ (Fig.~\ref{fig:activation_landscapes}e), a perversion of the expected behavior of the network, highlighting its failure to properly weigh and learn the output for all inputs equally in this configuration.

\begin{table}[!h]
\setlength{\tabcolsep}{7pt}
\begin{tabular}{cccc}
\hline
Learned function  & Sinusoidal activation & Sigmoid activation by RUS & Sigmoid activation w/o correction\\
\hline
\multicolumn{4}{c}{Minimum $\mC$} \\
\hline
NAND              & 0.269             & 0.330           & 0.357 \\
XOR               & 0.063             & 0.240           & 0.264\\
IMPLICATION2      & 0.273             & 0.370           & 0.373\\
\hline

\hline
\multicolumn{4}{c}{Parameters minimizing $\mC$ ($^{\circ}$)} \\
\hline
NAND              & (126,74,127)       & (82,108,132)     & (0,0,180)\\
XOR               & (164,146,359)       & (180,180,0)       & (136,132,43)\\
IMPLICATION2              & (100,284,208)      & (254,120,123)    & (240,86,163)\\
\hline
\end{tabular}
\caption{\textbf{Comparison of training performance for variations of the activation function circuit.} Comparison of minimum value of $\mC$ (and respective parameters) achieved after training with three different activation circuits: original gearbox (sigmoid-like activation), single iteration of the original circuit without correction on failure, and Rabi activation (sinusoidal-like activation) through the Rabi oscillation of the qubit.}
\label{tab:activation_comparison}
\end{table}

Having performed training using the same procedure for all three circuit variations, the results in both minimum cost-function value achieved and learned parameters are compiled (Table~\ref{tab:activation_comparison}) for a quantitative comparison. They show that in all instances, the value of $\mC$ is higher for the original activation circuit without correction than for the same circuit making full use of RUS, demonstrating the usefulness of this strategy already in this limited scenario. However, note that for the Rabi activation circuit, $\mC$ can be further minimized in all instances, owing to the severely reduced circuit depth of the circuits implementing this use case. Further improvements in the fidelity of operations and in the mitigation of parasitic interactions should limit the advantage of the Rabi activation circuit, as was indeed verified in simulation. Nevertheless, this realization represents an important lesson about the necessity of tailoring quantum applications to the characteristics of the hardware. Indeed, this suggests the possibility of using the Rabi oscillation of a qubit as an intrinsic (soft) non-linearity for optimized implementations of QNNs on near-term superconducting quantum processors.

\section{Simulation Methodology}
\label{si:sec:simulation-methodology}

We perform density-matrix simulations for an ideal processor and a noisy one using \textit{Quantumsim}~\cite{Obrien17}. Due to parameter drift, we prefer to set some parameters of the error model from a simultaneous best fit to experimental data. For this purpose we use the data from a relatively simple circuit, namely the single-pass, three-qubit conditional gearbox circuit of Fig.~\ref{fig:sigmoid_synthesis}. Unfortunately, we lack the equivalent of Fig.~\ref{fig:sigmoid_synthesis} for a three-qubit conditional gearbox circuit using $\QItwo$ instead of $\QIone$.
We list the error sources considered and the way we set their parameter values:
\begin{enumerate}
\item
Qubit relaxation and dephasing. We use the $\Tone$ and $\Ttwoecho$ values listed in Table~\ref{tab:device_performance}.
\item
Qubit initialization error due to residual excitation. We set the error for $\QA$, $\QIone$, and $\QO$ from a fit to Fig.~\ref{fig:sigmoid_synthesis}. 
We neglect initialization error on $\QItwo$.
\item
Misclassification of the $\QA$ measurement outcome. We use a fit to Fig.~\ref{fig:sigmoid_synthesis}.
\item Residual $ZZ$ coupling between $\QA$ and $\QO$. For simplicity, we do not include the much weaker $ZZ$ coupling for other qubit pairs.
We model this effect during single-qubit gates by adding an extra $\ket{11}\bra{11}$ term to the Hamiltonian and obtain the gate action via first-order Trotterization.
We use the calibrated $ZZ$ coupling strength of Fig.~\ref{fig:residual_zz}, using the average of the two corresponding non-diagonal entries.
\item
Cross-resonance during single-qubit gates acting on $\QA$. 
This effect arises from driving $\QA$ via the common feedline rather than through its dedicated microwave drive line. 
We model cross-resonance effects on $\QA$ only for $\QO$, the qubit with least detuning from $\QA$. 
Specifically, the actual rotation angle of $R_x$ and $R_y$ gates on $\QA$ is scaled by  $1+\alpha_k$, where $\alpha_k$ depends on the state of $\QO$. 
Here, we set $\alpha_0=0$ since single-qubit gates on $\QA$ are calibrated with $\QO$ in $\ket{0}$, and set $\alpha_1$ from a fit to Fig.~\ref{fig:sigmoid_synthesis}. 
\item
Remaining drive non-linearity of single-qubit gates acting on $\QA$ after the compensation of Fig.~\ref{fig:microwave_characterization}.
The overall transformation from the nominal rotation angle $\theta$ to the actual rotation angle $\theta_{\mathrm{eff}}$ is modelled by
\[
\theta_{\mathrm{eff}} = \pi \frac{\sin(\frac{\theta}{f_{\mathrm{NL}}})}{\sin(\frac{\pi}{f_{\mathrm{NL}}})},
\]
where $f_{\mathrm{NL}}$ is a non-linearity factor. This form captures the dominant third-order nonlinearity. The smaller the value of $f_{\mathrm{NL}}$, the stronger the non-linearity. We set $f_{\mathrm{NL}}$ from a fit to Fig.~\ref{fig:sigmoid_synthesis}.
\item
Coherent phase errors during CZ gates.  We simulate the phase action of each CZ gate as a four-qubit operation according to
Eq.~\eqref{eq:cz-model}, but truncating terms with negligible phase. In practice, the dominant phases errors are on terms $Z_A$, $Z_O$, and $Z_A Z_O$.
We obtain these errors for $\mathrm{CZ}(\QA,\QIone)$ and $\mathrm{CZ}(\QA,\QO)$ from a fit to Fig.~\ref{fig:sigmoid_synthesis}.
We do not include errors on $\mathrm{CZ}(\QA,\QItwo)$.
\item
Increased dephasing of flux-pulsed qubits during CZ gates. The higher-frequency transmon in the pair is pulsed away from the sweetspot. In addition, for $\mathrm{CZ}(\QA,\QIone)$ (respectively $\mathrm{CZ}(\QA,\QItwo)$), the spectator qubit $\QItwo$ (respectively $\QIone$) must also be pulsed away (this action is sometimes referred to as ``parking'').
All such pulsing causes a suppression of $\Ttwoecho$. For simplicity, we take $\Ttwoecho$ suppression to be the same for all pulsed qubits, setting the value from a fit to Fig.~\ref{fig:sigmoid_synthesis}. 
\item
Measurement-induced phase shift. Mid-circuit measurements of $\QA$ induce a phase on $\QO$ which is different depending on whether $\QA$ is collapsed to $\ket{0}$ or $\ket{1}$. We use the $0\degrees$ $(10\degrees)$ phase calibrated for $\QA$ collapse to $\ket{0}$ $(\ket{1})$.
\end{enumerate}

\begin{figure}[!b]
\centering
\includegraphics[width=0.3\textwidth]{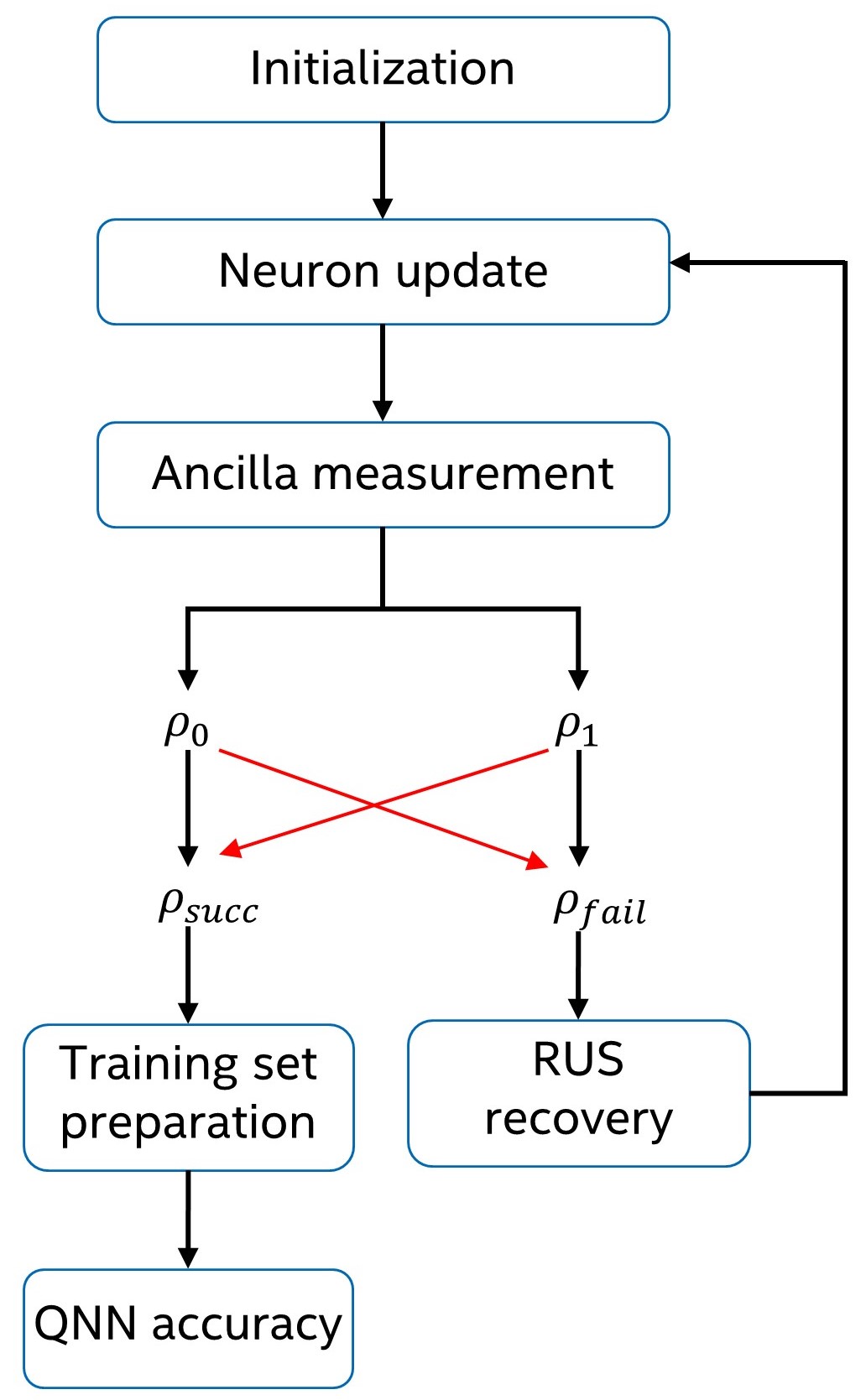}
\caption{
\textbf{Simulation technique to account for the stochasticity in repeat-until-sucess circuits.}
The red arrows indicate how we incorporate errors due to $\QA$ readout misclassification in the RUS procedure.
}
\label{fig:simulation_scheme}
\end{figure}

\begin{figure}[!b]
\centering
\includegraphics[width=0.63\textwidth]{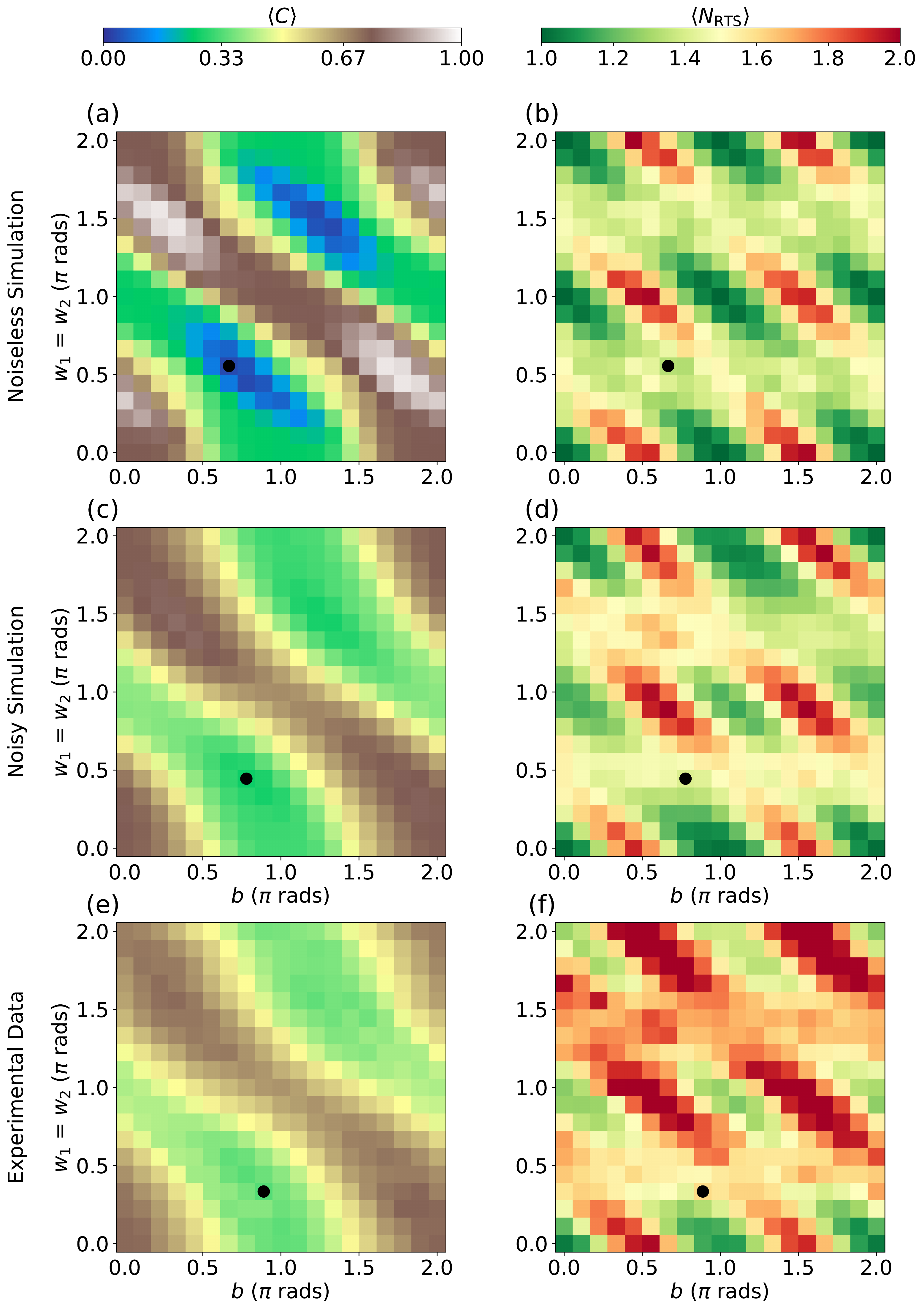}
\caption{
\textbf{Simulation of feature space landscapes for the NAND function.}
Top panels show noiseless simulation of (a) $\mC$ and (b) $\mN$ for NAND along 2-D slices of the 3-parameter space. Similarly, panels (c and d) show noisy simulation results and panels (e and f) the corresponding experimental results
(same as Figs.~\ref{fig:feature_spaces}e and~\ref{fig:feature_spaces}f, respectively), reproduced here to facilitate comparison.
Black dots indicate the parameters minimizing $\mC$ within the slice.
}
\label{fig:nand_simulation}
\end{figure}

The stochastic nature of the RUS-backed conditional gearbox circuit is taken into account in simulation in the following manner. The measurement of $\QA$ as part of the neuron update collapses the state of $\QA$ to one of two density matrices, $\rho_0$ and $\rho_1$, depending on the ancilla qubit collapsing to $\ket{0}$ or $\ket{1}$, respectively. Since the simulator maintains a complete representation of the quantum state at each point of the circuit, we have complete access to the two (un-normalized) density matrices.
We apply the measurement-induced phase to $\QO$. Then we apply the misclassification of the measurement outcome with probability $p$, leading to density matrices
\[
\begin{split}
\rho_{\mathrm{succ}} &= (1-p)\rho_0 + p \,\rho_1,\\
\rho_{\mathrm{fail}} &= (1-p) \rho_1 + p \, \rho_0
\end{split}
\]
corresponding to declared success and failure, respectively. At this point we apply the remaining circuit (parity-check comparison and training-set preparation) to $\rho_{\mathrm{succ}}$ and the correction sub-circuit and then repeat the neuron-update step for $\rho_{\mathrm{fail}}$.
The simulation results are obtained as the incoherent sum of the $\rho_{\mathrm{succ}}$ at each attempt. Notice that $\rho_{\mathrm{succ}}$ and $\rho_{\mathrm{fail}}$ are not normalized and that their norm represents the probability of the corresponding history of failures and success. Figure~\ref{fig:simulation_scheme} helps to visualize the method described.

\section{Simulation results}
\label{si:sec:simulation-results}

\begin{figure}[!ht]
\centering
\includegraphics[width=\textwidth]{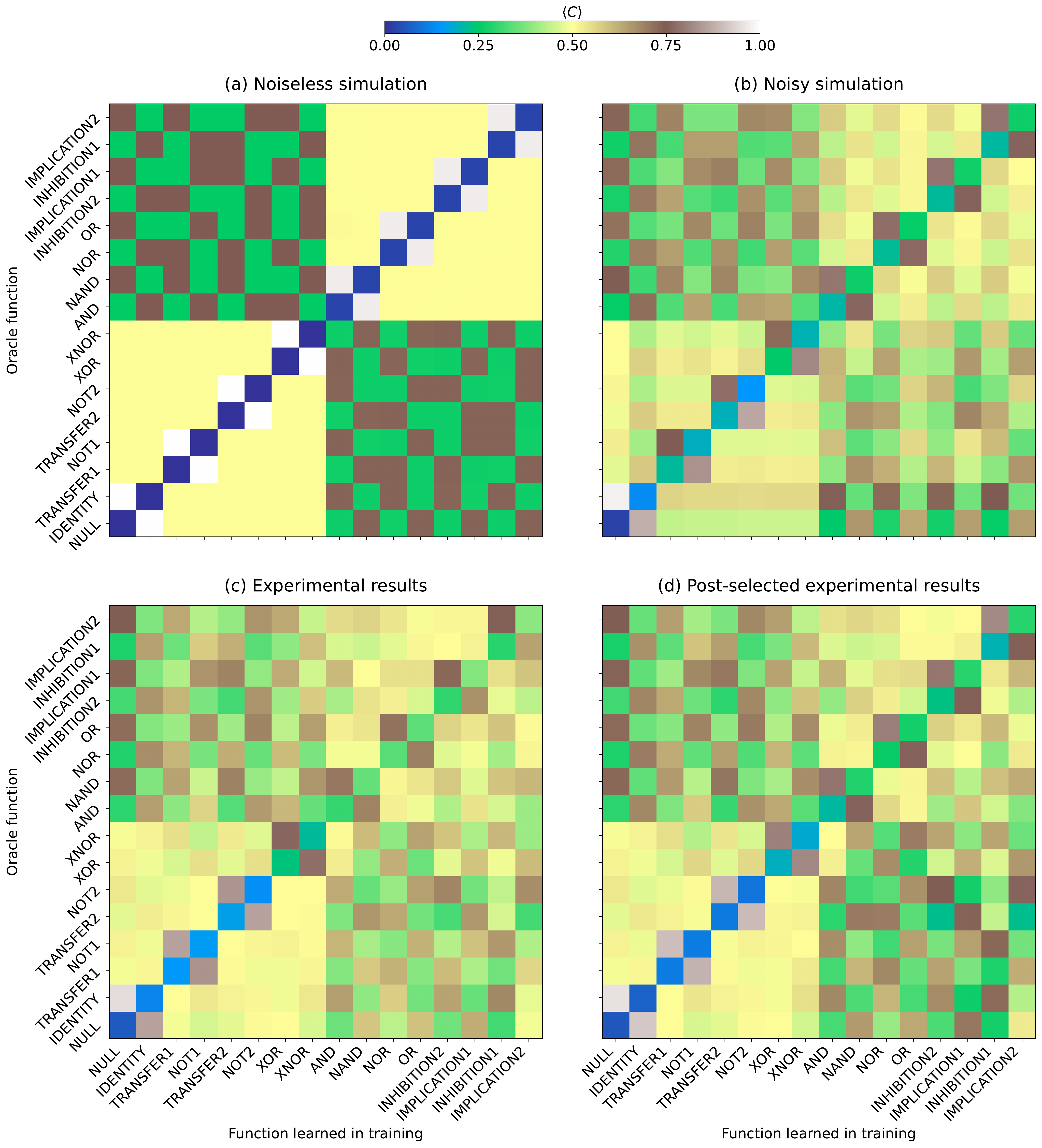}
\caption{
\textbf{Comparison of simulated and experimental specificity matrices.}
(a) Simulated specificity matrix for an ideal processor. There are 7 values for $\mC$ across all combinations of ideal parameters and oracle functions:
0 (dark blue), 0.25 (green), 0.5 (yellow), 0.75 (brown), and 1 (white).
(b) Noisy simulation of the specificity matrix, with optimal parameters learned by training in noisy simulation.
(c) Experimental specificity matrix. This is the same matrix as shown in Fig.~\ref{fig:specificity_matrix}, reproduced here to facilitate comparison.
(d) Experimental specificity matrix obtained by post-selecting the raw data in (c) on success in the first iteration of the conditional gearbox circuit.
The higher contrast in (d) relative to (c) is due to the minimized circuit depth of the threshold activation under this post-selection condition.
}
\label{fig:specificity_matrices}
\end{figure}

In this section we provide simulation results for an ideal processor and a noisy one, and compare them to experimental data. We show the feature landscapes for NAND in Fig.~\ref{fig:nand_simulation}. The noisy simulation qualitatively matches the experiment, with similar distortion and reduced contrast of the feature space landscape relative to the ideal simulation. Quantitative discrepancies between noisy simulation and experiment likely result form additional errors not included in simulation, most notably transmon leakage during CZ gates. 
Note that the minimal $\mC$, indicated by the black dot in the left panels, is achieved for different values of $(w_1, w_2, b)$.
For an ideal processor, there are multiple $(w_1, w_2, b)$ that globally minimize $\mC$, due to symmetry. 
Error breaks the symmetry both in noisy simulation and experiment. 
Finally,  we compare the specificity matrices obtained in simulation and experiment (Fig.~\ref{fig:specificity_matrices}).
The horizontal axis corresponds to the Boolean function used in training, while the vertical axis corresponds to the Boolean function used to test the network.

We use ideal simulation to gain familiarity with the expected structure of the specificity matrix.
For optimal parameters $(w_1, w_2, b)$, we use any one of the several choices that minimize $\mC$.
The lowest values of $\mC$ are found along the diagonal. This is expected, as in this case learning and testing functions match.
Constant and balanced functions can be perfectly learned, and thus $\mC=0$ for these.
Unbalanced functions cannot be perfectly learned, and for these we find $\mC\approx0.029$.
Half of the next-to-diagonal elements have $\mC$ equal to or close to unity, since the testing function is the complement of the function learned (the specific function whose output differs for all four inputs).
For all entries, $\mC$ is equal to or close to a multiple of 0.25, the multiple corresponding to the fraction of input states for which the output of the learned and testing functions differs.
For any two constant of balanced functions, the outputs differ for 0, 2, or 4 inputs. The same holds for any two unbalanced functions. This explains the structure of the lower-left and top-right quadrants. Outputs for any constant/balanced function differ from those of any balanced function for either 1 or 3 inputs. This explains the structure of the top-left and bottom-right quadrants.

For noisy simulation we use the optimal parameters obtained in simulated training.
We observe a qualitative similarity between experiment and noisy simulation.


\end{bibunit}

\end{document}